\documentclass[universe,review,accept,oneauthor,pdftex,aasmacros]{Definitions/mdpi} 

\firstpage{1} 
\pubvolume{8}
\issuenum{8}
\articlenumber{392}
\pubyear{2022}
\copyrightyear{2022}
\externaleditor{Academic Editors: Anna Sajina and Asantha R. Cooray} 
\datereceived{25 June 2022} 
\dateaccepted{20 July 2022} 
\datepublished{26 July 2022} 
\hreflink{https://doi.org/10.3390/universe8080392} 
\pdfoutput=1

\newcommand{\feii}{[Fe \textsc{ii}]}
\newcommand{\fevii}{[Fe \textsc{vii}]}
\newcommand{\paa}{Pa$\alpha$}
\newcommand{\pab}{Pa$\beta$}
\newcommand{\molhy}{H$_2$}
\newcommand{\brg}{Br$\gamma$}
\newcommand{\brd}{Br$\delta$}
\newcommand{\ha}{H$\alpha$}
\newcommand{\hb}{H$\beta$}

\newcommand{\ovi}{[O \textsc{vi}]}
\newcommand{\oiv}{[O \textsc{iv}]}
\newcommand{\oiii}{[O \textsc{iii}]}
\newcommand{\oii}{[O \textsc{ii}]}

\newcommand{\nii}{[N \textsc{ii}]}
\newcommand{\heii}{He \textsc{ii}}
\newcommand{\neii}{[Ne \textsc{ii}]}
\newcommand{\neiii}{[Ne \textsc{iii}]}
\newcommand{\nev}{[Ne \textsc{v}]}
\newcommand{\nevi}{[Ne \textsc{vi}]}
\newcommand{\nv}{[N \textsc{v}]}
\newcommand{\sii}{[S \textsc{ii}]}

\newcommand{\sivi}{[Si \textsc{vi}]}
\newcommand{\mgiv}{[Mg \textsc{iv}]}
\newcommand{\arvi}{[Ar \textsc{vi}]}
\newcommand{\farcs}{\mbox{\ensuremath{.\!\!^{\prime\prime}}}}%
\newcommand{\arcsec}{\mbox{\ensuremath{^{\prime\prime}}}}%

\Title{The Role of AGN in Luminous Infrared Galaxies from the Multiwavelength Perspective}

\TitleCitation{The Role of AGN in Luminous Infrared Galaxies from the Multiwavelength Perspective}

\Author{Vivian U \orcidA{}} 

\AuthorNames{U, Vivian}

\AuthorCitation{U, V.}

\address[1]{%
Department of Physics and Astronomy, University of California, 
Irvine, CA 92697, USA; vivianu@uci.edu} 

\abstract{Galaxy mergers provide a mechanism for galaxies to effectively funnel gas and materials toward their nuclei and fuel the central starbursts and accretion of supermassive black holes. In turn, the active nuclei drive galactic-scale outflows that subsequently impact the evolution of the host galaxies. The details of this transformative process as they pertain to the supermassive black holes remain ambiguous, partially due to the central obscuration commonly found in the dust-reddened merger hosts, and also because there are relatively few laboratories in the nearby universe where the process can be studied in depth. This review highlights the current state of the literature on the role of accreting supermassive black holes in local luminous infrared galaxies as seen from  various windows within the electromagnetic spectrum. Specifically, we discuss the multiwavelength signatures of the active nucleus, its associated feeding and feedback processes, and the implications of multiple supermassive black holes found in nearby interacting galaxy systems for galaxy evolution from the observational perspective. We conclude with a future outlook on how the topic of active nuclei in low- and high-redshift galaxy mergers will benefit from the advent of next-generation observing facilities with unparalleled resolving power and sensitivity in the coming decade.
}

\keyword{Luminous infrared galaxies; Galaxy mergers; Seyfert galaxies; Starburst galaxies; Active galactic nuclei; Supermassive black holes; Galaxy interactions; Kinematics and dynamics}  

\begin{document}

\section{Introduction}

Galaxies, the fundamental building blocks of the universe, form and evolve over cosmic time. They might simply have cold gas accretion from the cosmic web onto their galactic disks to form stars~\citep{Martin16}. However, the $\Lambda$ cold dark matter (CDM) cosmological model dictates that structure grows hierarchically~\citep{White78,Blumenthal84}, so merging seems to be an inevitable stage within a massive galaxy's lifetime.  Often in field or loose group environments, spiral galaxies encounter a near-equal mass companion with which they engage in a major merger. Such collision events bring about dramatic changes to their stellar content, gas content, central supermassive black hole (SMBH), and their overall fate, as stars and gas lose angular momentum and funnel toward the dynamical center~\citep{Sanders88,hopkins_cosmological_2008}. How the properties of these individual components are affected by the encounter depends on the specific circumstances associated with the interaction, but one common phase for these colliding systems is when they display extreme infrared excess from dust emission and high rates of massive star formation~\citep{Rieke79,Lonsdale84}. This infrared excess leads to an observable bias, dubbing this population of galaxies---``infrared galaxies''. 

The study of infrared galaxies surged with the launch of the Infrared Astronomical Satellite (\emph{IRAS})~\citep{Neugebauer84} in 1983, which enabled the first unbiased, sensitive, all-sky infrared survey~\citep{Beichman88}. This survey brought about, among an assortment of catalogs and atlases, the \emph{IRAS} Bright Galaxy Sample~\citep{Soifer87}, a complete sample of the brightest, far-infrared 60-$\upmu$m~selected galaxies. Over the past three decades, luminous infrared galaxies (LIRGs, with the empirical definition $L_\mathrm{IR}$[8--1000 $\upmu$m] $\geq$ 10$^{11} L_\odot$) and their ultraluminous counterparts (ULIRGs, with $L_\mathrm{IR} \geq 10^{12} L_\odot$) 
 have been studied extensively. Comparing their luminosity functions to those of normal galaxies~\citep{Schechter76}, starbursts, Seyferts~\citep{Huchra77}, and Palomar--Green quasars~\citep{Schmidt83} reveals that LIRGs dominate luminosities above $\sim$$2 \times 10^{11} L_\odot$ in the local universe, matching the space densities of Seyferts at the lower luminosities and surpassing those of quasars at the high end~\citep{Soifer87,sanders_luminous_1996}. The bolometric luminosities and space densities of ULIRGs are found to be at least a factor of $\sim$1.5--2 times larger than those of the optically-selected quasi-stellar objects~(QSOs; \cite{Kim98,Surace98}), though they are dynamically similar~\citep{Rothberg13}.

Redshift surveys conducted using the \emph{Infrared Space Observatory}~\citep{Kessler96}, SCUBA on the James Clerk Maxwell Telescope~\citep{Holland99}, and the {\emph{Spitzer Space Telescope}}~\citep{Werner04} find very strong evolution in the luminosity function of (U)LIRGs~\citep{lefloch05}, dominating star formation activity beyond redshifts $z > 0.7$~\citep{Casey14}. Given their prominence at the peak of cosmic history ($z \sim 2$), and that a substantial fraction of galaxies pass through this intense infrared emission stage~\citep{Soifer87}, (U)LIRGs are believed to play a significant role in driving the cosmic star formation rate (SFR) density with which their cosmic histories coincide~\citep{Madau14}. Results from high-redshift (U)LIRG studies conducted using deep surveys, such as the Cosmic Evolution Survey~\citep{Scoville07}, the Great Observatories Origins Deep Survey~\citep{Elbaz11}, and the Cosmic Assembly Near-Infrared Dark Energy Legacy Survey~\citep{Grogin11,Koekemoer11} suggest that interactions and infrared luminosities are intimately linked~\citep{Zheng04,Bell05,Bridge07,Kartaltepe10,Kartaltepe12}. This is particularly true at low redshifts where all of the ULIRGs, and many of the LIRGs, are gas-rich major mergers~\citep{Sanders88,Larson20}.

\begin{figure}[H]
  \includegraphics[width=0.75\textwidth]{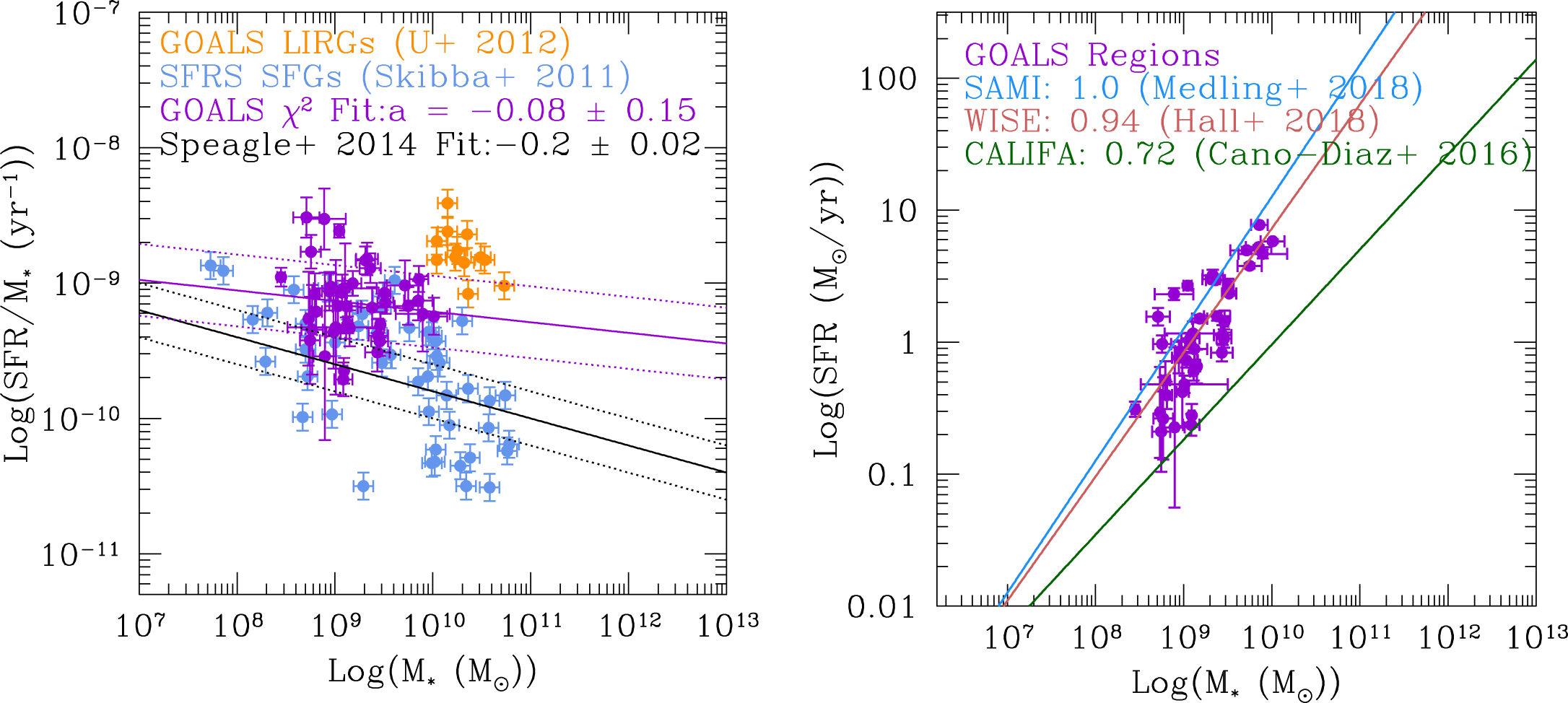}
  \caption{(\textbf{Left}) The specific SFR (=SFR/M$_*$) distribution as a function of the stellar mass shows that the galaxy-integrated (orange) and extranuclear star-forming regions (purple) in GOALS LIRGs are above local star-forming galaxies (blue; \cite{Skibba11}) or the global fit from~Speagle et~al.~\cite{Speagle14}. (\textbf{Right}) The SFMS for the GOALS star-forming regions plotted against the normalized resolved galaxy main sequence from other surveys~\citep{Cano-Diaz16,Hall18,Medling18}. Figure from~Linden et~al.~\cite{Linden19};~\textcopyright~AAS, reproduced with permission.
  \label{fig:sfms}} 
\end{figure}

While it is clear that (U)LIRGs are an important population of objects to understand given their place in cosmic history, whether galaxies of similar luminosity at high redshifts are analogs of those locally is much more nuanced. If low-$z$ and high-$z$ (U)LIRGs are alike, what physical processes connect these two populations given that galaxies evolve? The fundamental link between the nearby objects and their distant cousins may not be the absolute value of the infrared luminosity but the amount of gas content, which dictates SFR and feedback efficiencies for these galaxies that lie above the star formation main sequence (SFMS) (see Figure~\ref{fig:sfms}; \cite{Wuyts11,Speagle14,Linden19}).
Thus, it is pertinent to conduct in-depth studies of the physical processes taking place within local (U)LIRGs where we can resolve details to appreciate how gas ultimately drives galaxy evolution through the merger mechanism. 

The motivation to understand the nature of (U)LIRGs and their role in galaxy evolution prompted the Great Observatories All-sky LIRGs Survey~(GOALS; \cite{armus_goals_2009}). Comprising the 200 brightest 60 $\upmu$m-selected (U)LIRGs in the nearby ($z \leq 0.08$) universe, the GOALS sample stemmed from the \emph{IRAS} Revised Bright Galaxy Sample~\citep{sanders_iras_2003} that includes 629 extragalactic objects with $f_{60}$ $>$ 5.24 Jy at galactic latitudes $\vert b \vert > 5^\circ$. The data repository for GOALS is extensive across the electromagnetic spectrum and is growing, encompassing data taken by the \emph{Nuclear Spectroscopic Telescope Array} (\emph{NuSTAR}), the \emph{Chandra X-ray Observatory}, the \emph{Galaxy Evolution Explorer} (\emph{GALEX}), the \emph{Hubble Space Telescope} (\emph{HST}), the Keck Telescopes, the Very Large Telescopes, the \emph{Spitzer Space Telescope}, the Atacama Large Millimeter/submillimeter Array (ALMA), the Karl G. Jansky Very Large Array (VLA), and imminently, the \emph{James Webb Space Telescope} (\emph{JWST}). The
rich ancillary data set has facilitated a large number of studies on the properties of star formation,the interstellar medium (ISM), dust, and SMBHs in (U)LIRGs and enriched the literature with results on star cluster formation and destruction, ionization and shocks, nuclear obscuration, and others. This review highlights many of the SMBH-related results from GOALS as well as  other non-GOALS (U)LIRG studies. 


Complementing other articles in this Special Issue---``Infrared Spectral Energy Distribution and Variability of Active Galactic Nuclei: Clues to the Structure of Circumnuclear Material'' (Lyu and Rieke 2022~\citep{Lyu22}), ``The past and future of mid-infrared studies of AGN'' (Sajina, Lacy, and Pope 2022~\citep{Sajina22}), ``Molecular gas heating, star-formation rate relations and AGN feedback in Ultraluminous Infrared Galaxies'' (Farrah 2022), among others---this review focuses on the coevolution of SMBHs and galaxies. In particular, we emphasize the role the central SMBH (often found in an actively accreting phase as an active galactic nucleus--nuclei (AGN) during the late stages of merging) plays in influencing its host galaxy. An outline is laid out as follows: In Section~\ref{sec:agn}, we present the conditions of the inner kiloparsec environment and the nature of the nuclear energy source in infrared galaxies as viewed from the multiwavelength perspective. We then discuss the interplay between the AGN and the circumnuclear interstellar medium (ISM) and beyond in Section~\ref{sec:feedback}. In Section~\ref{sec:duals}, the implications of the multiple interacting systems and the significance of the merging population in the multi-messenger era are discussed. Finally, the future observational outlook with selected telescope facilities is summarized in Section~\ref{sec:future} while the key conclusions are given in Section~\ref{sec:conclusions}. 

\section{The Central Regions of (U)LIRGs}
\label{sec:agn}

\begin{figure}[H]

    \begin{minipage}{0.40\textwidth}
     \includegraphics[width=1.3\textwidth]{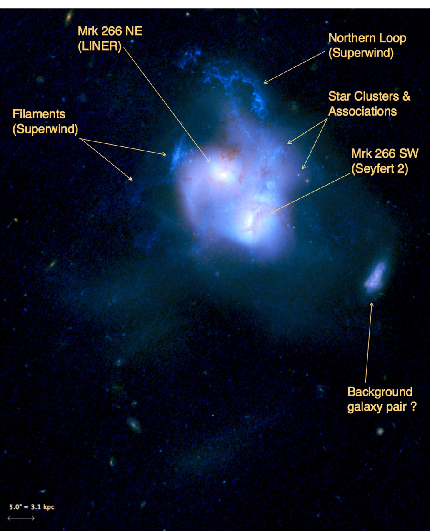}
  \end{minipage}\hspace{1.8cm}
  \begin{minipage}{0.31\textwidth}
  \includegraphics[width=1.4\textwidth]{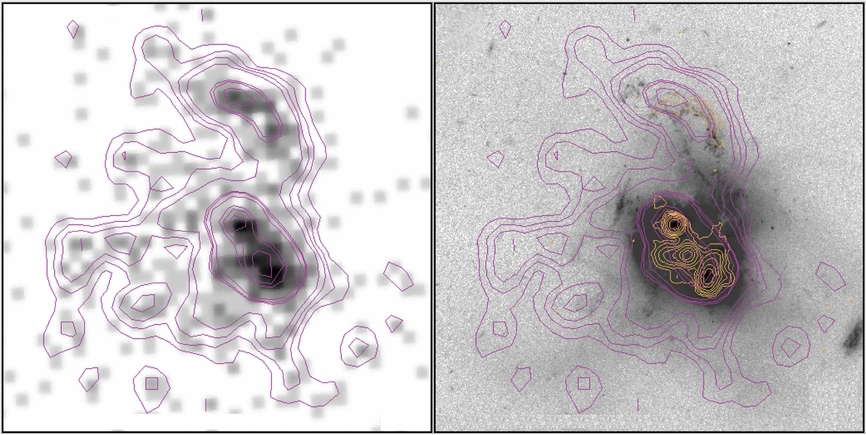}
  \includegraphics[width=1.4\textwidth]{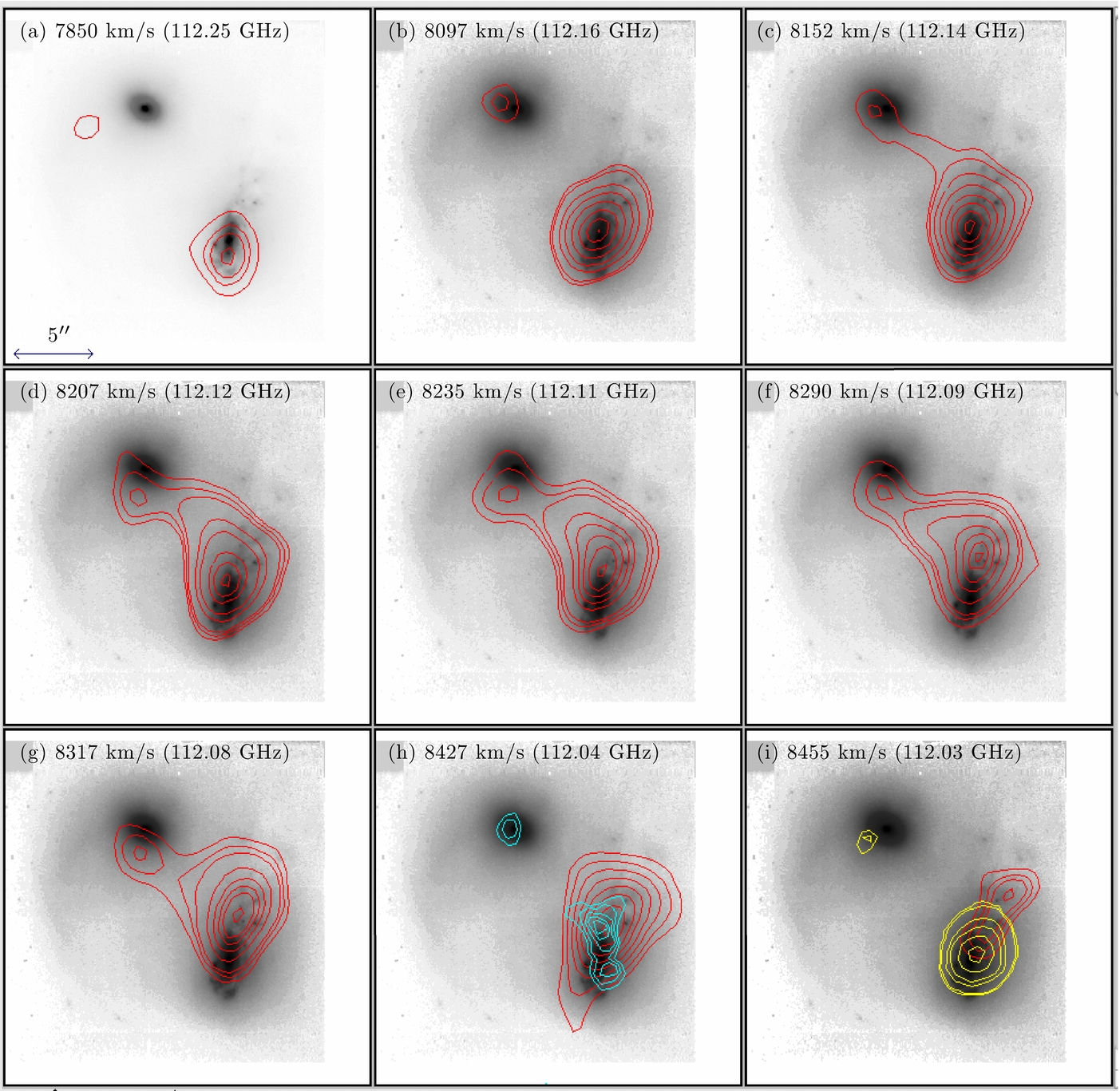}
  \end{minipage}
  \caption{Mrk 
266 in multiwavelength light---(\textbf{Left}) \emph{HST} composite image showing major structures; (\textbf{Right, Top}) soft X-ray contours (magenta contours) compared to the \emph{HST} $F435W$-band image and VLA radio continuum (orange contours); (\textbf{Right, Bottom}) CO ($1-0$) channel maps (red contours), integrated HCN ($1-0$) emission (cyan contours), and integrated HCO$^+$ ($1-0$) emission (yellow contours) overlaid on \emph{HST} 1.6 $\upmu$m images. This dual AGN system exhibits an earlier dust blow-out phase and more rapid gas consumption than predicted from simulations, revealing insight into the evolutionary timeline of large-scale superwinds. This study illustrates the complexity of dust and gas kinematics that can only be teased apart with detailed analyses of high-resolution multiwavelength data. Figure adapted from~Mazzarella et~al.~\cite{Mazzarella12};~\textcopyright~AAS, reproduced with permission.} 
  \label{fig:mrk266}
\end{figure}

\begin{figure}[H]

\includegraphics[width=.6\textwidth]{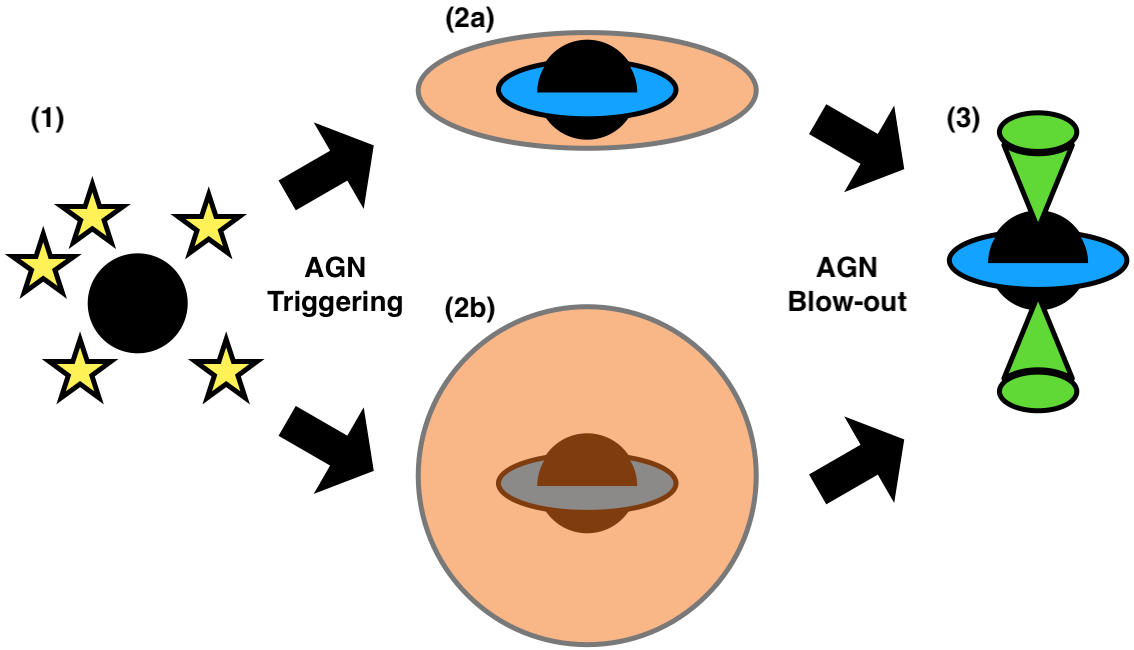}
\caption{Schematic of plausible black hole evolutionary paths along the galaxy merger sequence. (\textbf{1}) In early-stage mergers, the SMBH is quiescent and the nuclear activity is dominated by star formation. As interaction proceeds and gas is funneled toward the accretion disk of the SMBH, the AGN phase is triggered---though it may be (\textbf{2a}) a low-luminosity AGN partially obscured by a dusty torus or (\textbf{2b}) a fully buried Compton Thick AGN. (\textbf{3}) In the blow-out phase, the AGN luminosity may then dominate the nuclear energetics. It is unclear at which time point one or more of the SMBHs in merging progenitors achieve these different stages, which likely depend on the local conditions of the ISM as well as the geometry of the encounter or flickering of the central engine due to the AGN duty cycle. Putting together multiwavelength data sets for a statistical sample of galaxy systems at different merger stages will help constrain the simultaneity of AGN-triggering and shed insight into dual AGN fractions on merger timescales. 
\label{fig:bhpath}} 
\end{figure}

Although infrared galaxies are characterized by their infrared excess, or characteristic infrared bump peaking at $\sim$100~$\upmu$m~\citep{sanders_luminous_1996,u_spectral_2012,Paspaliaris21} due to the large dust component~\citep{Surace98}, they are multiwavelength objects by nature---emission from the various subcomponents of the galaxy system radiates in different regions of the electromagnetic spectrum (Figure~\ref{fig:mrk266}). As a result, locating the true dynamical center where the SMBH resides is nontrivial---particularly for interacting mergers engaged in a closely-separated pair with clearly distorted host morphology. The challenge arises partly because the relative contribution to the total luminosity from the SMBH versus star formation is difficult to distinguish at resolved scales (e.g., dusty star-forming regions or AGN dusty torus)~\citep{RamosPadilla20}, and partially because many nuclei appear clumpy~\citep{medling_stellar_2014,Larson20,Linden21}. Furthermore, the nature of the SMBH is difficult to discern in these transitioning systems: quiescent SMBHs may be found in the early stages of mergers before gas accretion into the central region; deeply buried AGN~\citep{Imanishi08,Imanishi10,Lee12} may be expected in mid-stage mergers where the SMBH is actively accreting gas but hidden behind a thick cocoon of dust, or actively accreting AGN that has emerged after blowing out the dust through radiation pressure in the late stages of merging (Figure~\ref{fig:bhpath}). At any time point, one or both of the SMBHs might be experiencing these different states. Thus, multiwavelength diagnostic tools have been employed to verify the presence of AGN in (U)LIRGs. Here, we provide a broad overview of some recent observational results for AGN identification and characterization among local (U)LIRGs.

\vspace{-6pt}

\subsection{High-Energy Observations in the X-ray}
X-ray observations are regarded as robust tools for identifying AGN because the energetic photons from AGN can penetrate gas and dust more easily and are better distinguished from those of stars~\citep{Teng10,iwasawa_c-goals_2011,ricci_growing_2017}. This is a particularly notable attribute since (U)LIRGs are known to host high levels of dust extinction in the nuclei where AGN are likely to reside. Analyzing \emph{Chandra's} observations of 44 (U)LIRGs with $\log (L_\mathrm{IR}/L_\odot) = 11.73 - 12.50$, Iwasawa et~al.~\cite{iwasawa_c-goals_2011} identified AGN in 37 $\pm$ 6\% of the sample using an X-ray hardness ratio ($HR = (H-S)/(H+S) > -0.3$; where $H$ and $S$ are counts in the 2$-$8 keV and 0.5$-$2 keV bands, respectively~\citep{iwasawa09}) and the presence of the Fe K line at 6.4 keV~\citep{iwasawa_asca_1998,Vignati99,Ikebe00,Komossa03,Imanishi03,armus_goals_2009}. When inspecting additional \emph{Chandra} data of similar depths for lower-luminosity LIRGs, Torres-Alb{\`a} et al.~\cite{TorresAlba18} found a lower AGN fraction (31 $\pm$ 5\%) using X-ray and mid-infrared criteria, suggesting that AGN accretion grows with infrared luminosity. Intriguingly, Pereira-Santaella et~al.~\cite{Pereira-Santaella11} explored \emph{Chandra} and \emph{XMM-Newton} observations of 27 low-luminosity LIRGs and determined that $\sim$15\% of the non-Seyfert LIRGs had excessive hard-X-ray emissions relative to those expected from star formation, suggesting the presence of an obscured AGN. This result has interesting implications for the timescale of when the obscured AGN phase begins in the nuclei of (U)LIRGs.

Torres-Alb{\`a} et al.~\cite{TorresAlba18} also found that the fraction of dual AGN in systems that host at least one AGN is 29 $\pm$ 14\% (2 of 7), nearly three times higher than the 11 $\pm$ 10\% (1 of 9) found for the higher-luminosity sample. While these statistics are limited by the small sample sizes, the dual AGN fraction of the high-luminosity sample~\citep{iwasawa_c-goals_2011} falls short of previous theoretical~\citep{Capelo17} and observational X-rays selected~\citep{Koss12} and other merger studies~\citep{Ellison11,Satyapal17}.  The diminished dual AGN fraction among late-stage mergers may be partially explained by heavy obscuration given the insufficient sensitivity of the \emph{Chandra} data to detect the Fe K line, or its inability to resolve AGN pairs at separations $d_\mathrm{sep} \lesssim 200-300$ pc. This result has strong implications for the timescale of AGN triggering as well as merging the nuclei in the case of progenitor gravitational wave sources that will be further explored in Section~\ref{sec:duals}.

\begin{figure}[H]

\includegraphics[width=.49\textwidth]{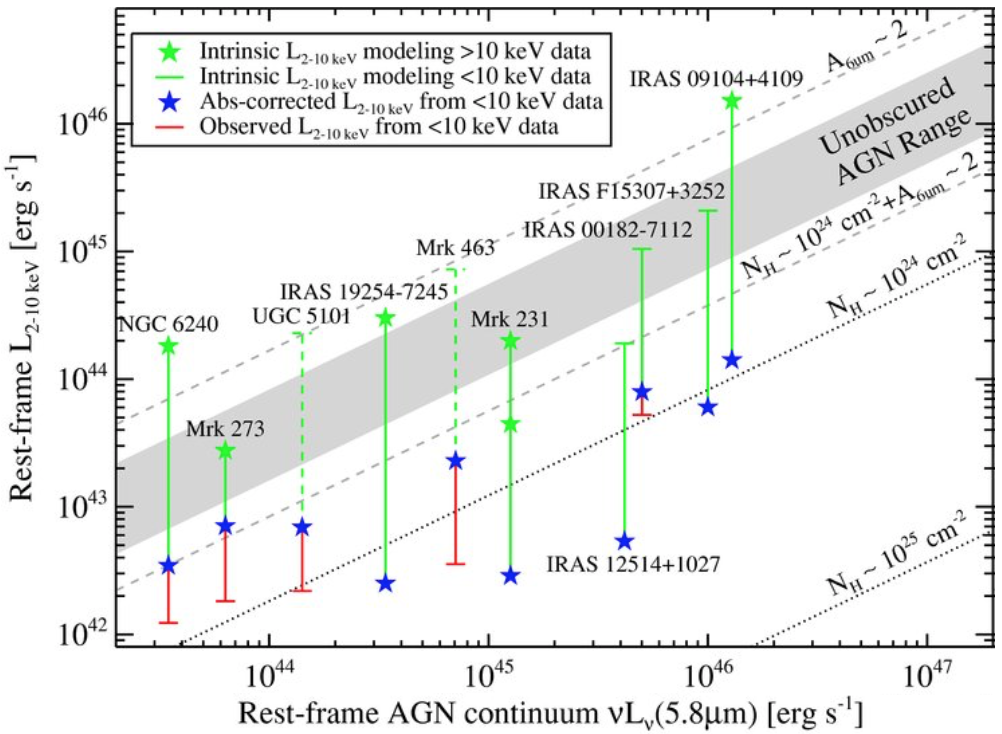}
\includegraphics[width=.49\textwidth,trim={0 390 0 0},clip]{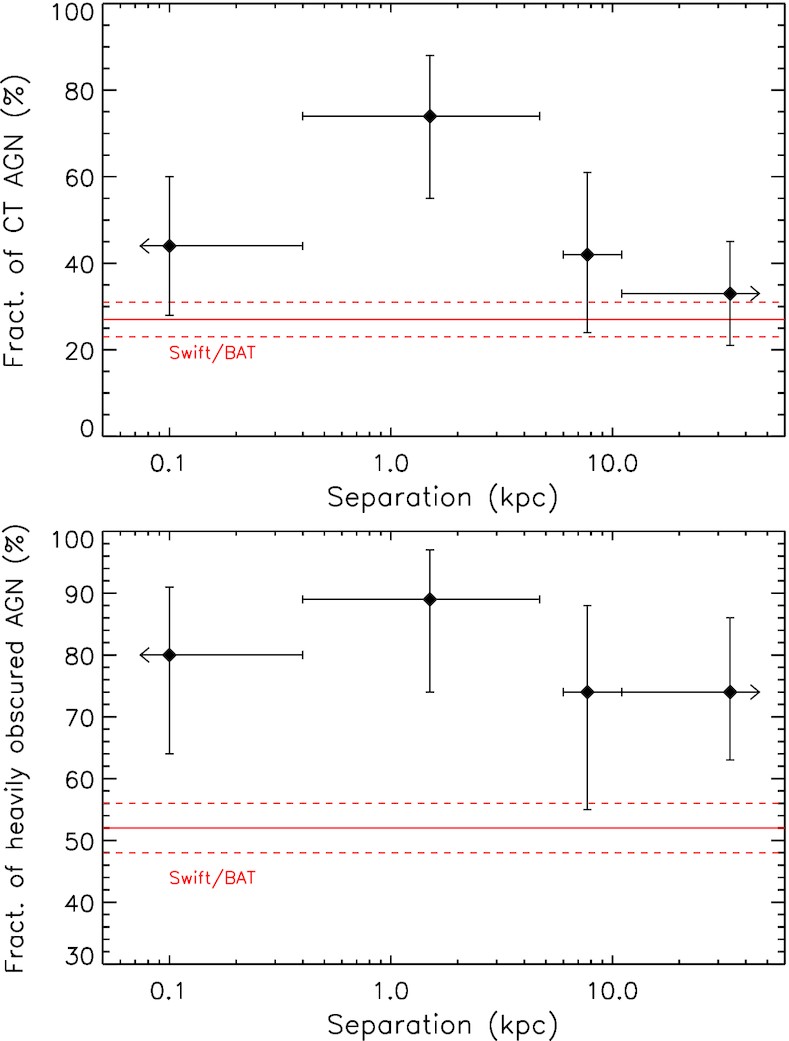}
\caption{(\textbf{Left}) The 
rest frame 2$-$10 keV luminosity versus 5.8 $\upmu$m~AGN continuum luminosity for several CT (or nearly CT) AGN-hosting ULIRGs. The gray shaded region indicates the range of unobscured AGN based on Type-1 sources, while the black dotted (dashed) lines demonstrate the effects of X-ray absorption by column densities of $N_H$ (with mid-infrared extinction by $A_{6 \mu m}$). The observed X-ray luminosities for the illustrated local ULIRGs residing around the $N_H \sim 10^{24}$~cm$^{-2}$ line lend evidence that ULIRGs are in fact highly obscured. Figure adapted from Bauer et~al.~\cite{Bauer10};~\textcopyright~AAS, reproduced with permission. (\textbf{Right}) Fraction of CT AGN as a function of projected nuclear separations with a peak found at a separation of a few kpc. The red continuous line marks the fraction of heavily obscured sources among \emph{Swift}/BAT AGN~\citep{Ricci15,ricci_bat_2017}. The resolution limit at which two nuclei could be separated at the distance of the depicted sample is $d_\mathrm{sep} \sim$ 0.4 kpc. Figure adapted from \mbox{Ricci et~al.~\cite{Ricci21}}.
\label{fig:ctagn}} 
\end{figure}

The launch of the hard X-ray observatory \emph{NuSTAR}~\citep{Harrison13} offers a way to locate the most obscured Compton Thick (CT, $N_\mathrm{H} \geq 10^{24}$ cm) sources buried in late-stage mergers, which turn out to be 65$^{+12}_{-13}$\% among AGN-hosting (U)LIRGs~\citep{ricci_growing_2017}. These ULIRGs lie along the column density $N_H \sim 10^{24}$ cm$^{-2}$ track on the Lutz et~al.~\cite{Lutz04} relation, which is a well-established empirical correlation between 2$-$10 keV and 5.8 $\upmu$m~luminosity for both Type-1 and absorption-corrected Type-2 AGN~\citep{Bauer10} (Figure~\ref{fig:ctagn}, left). A recent statistical \emph{NuSTAR} study of 60 (U)LIRGs showed that the fraction of CT AGN peaks at 74$^{+14}_{-19}$\% at small nuclear separations ($d_\mathrm{sep} \sim$ 0.4--6 kpc; \citep{Ricci21}, see also Figure~\ref{fig:ctagn}, right). This fraction is consistent with results from recent simulations showing that much of AGN activity remains undetected even in late-stage mergers~\citep{Blecha18}, 
despite using common mid-infrared selection techniques meant to overcome heavy obscuration. There is growing evidence that the obscuration properties of AGN in mergers differ from those of AGN in isolated galaxies; see recent reviews by \citet{RamosAlmeida17} and \citet{hickox_obscured_2018}. High-resolution multiwavelength tracers measuring the properties of the obscuring materials at resolved scales will be needed to assess the nuclear properties of the extraordinary nuclear environments in late-stage mergers (Gao, U {et al.}, in preparation).

\subsection{Spectroscopic Diagnostics Using Optical and Near-Infrared Emission Lines}
There are several widely-adopted optical diagnostics for detecting and characterizing AGN using spectroscopic measurements: the Baldwin, Phillips, and Terlevich~(BPT hereafter; \cite{Baldwin81})/Veilleux-Osterbrock~\citep{Veilleux87} diagrams based on excitation levels deduced from emission line ratios (see the in-depth discussion in the recent review by Kewley et~al.~\cite{kewley_understanding_2019}), and the Seyfert classifications based on the line profiles and widths of the hydrogen recombination lines. The BPT diagrams segregate galaxies or regions within galaxies by their energy source: theoretical~\citep{Kewley01} and empirical~\citep{Kauffmann03} limits for H \textsc{II} regions as well as the low-ionization nuclear emission-line region (LINER) 
and Seyfert division established by Kewley et~al.~\cite{Kewley01,Kewley06} and Kauffmann et~al.~\cite{Kauffmann03}. Applying this classification scheme to the 1 Jy ULIRG sample~\citep{Veilleux99} and LIRGs in the Bright Galaxy Survey sample~\citep{Veilleux95}, Yuan et~al.~\cite{Yuan10} found that there is a trend in the increasing ratio of Seyfert nuclei to H \textsc{II} regions with increasing infrared luminosity. The composite region between starburst- and AGN-dominated ionization was largely interpreted as being powered by a mix of star formation and low-luminosity AGN, but Rich et~al.~\cite{Rich11} studied two late-stage merging LIRGs with wide-field integral-field spectroscopy (IFS) and found that the composite emission can be explained by a combination of star formation and shocks without invoking AGN. 

\begin{figure}[H]

\includegraphics[width=.55\textwidth]{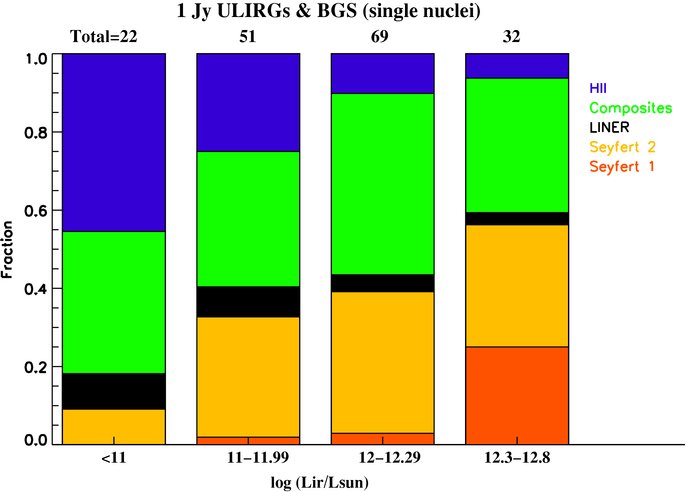}
\includegraphics[width=.38\textwidth]{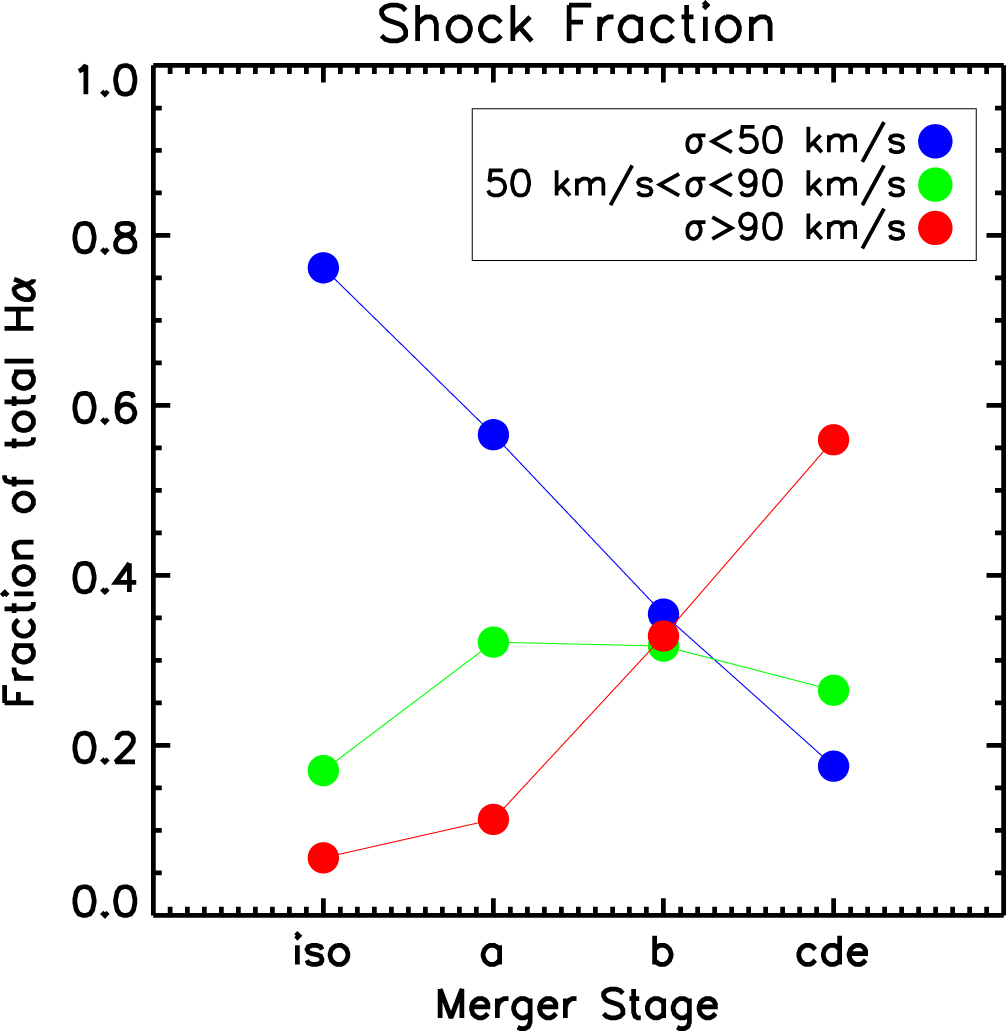}
\caption{(\textbf{Left}) BPT classification of the spectral type as a function of infrared luminosity for (U)LIRG samples from~Veilleux et~al.~\cite{Veilleux95,Veilleux99}. Seyfert nuclei dominate over H \textsc{II} regions as $L_\mathrm{IR}$ increases. Figure from Yuan et~al.~\cite{Yuan10}. (\textbf{Right}) The contribution from shocks to the \ha~luminosity is plotted against the merger stage, from isolated galaxies to early-stage and late-stage mergers. The data points are color-coded by the~\ha~velocity dispersion, where the dispersion cutoffs correspond to H \textsc{II} region velocity dispersions on the low end, and more turbulent and shock-dominated dispersions on the high end. The figure from Rich et~al.~\cite{Rich15}. Both panels~\textcopyright~AAS, reproduced with permission.
\label{fig:shocks}} 
\end{figure}

\textls[-5]{Investigating a large, representative sample of 27 (U)LIRGs in GOALS, \mbox{Rich et~al.~\cite{Rich14,Rich15}}} concluded that galaxy interaction--induced tidal forces result in galaxy-scale shock excitations (Figure~\ref{fig:shocks}).  
Shocks constitute a non-negligible fraction of the \ha~luminosity and may potentially be responsible for compressing the molecular gas and triggering starbursts at shock fronts throughout the merger.
An optical IFU study of 11 lower-luminosity LIRGs found large regions of enhanced \nii/\ha~and \sii/\ha~ratios associated with low surface-brightness H \textsc{ii} regions, with diffuse emission detected in \ha~and \paa~\citep{Alonso-Herrero09,Alonso-Herrero10}. There are likely mechanisms apart from young stars that are ionizing the gas, and these regions of enhanced low-ionization emission can be associated with the AGN ionization cone~\citep{Garcia-Marin06}.
While the prominence of galaxy-scale shocks is elucidated, the number of Seyfert nuclei identified using optical spectroscopic methods is likely a lower limit because infrared galaxies have dusty cores that would mask a broad feature in \ha~and \hb, or masquerade as dusty star-forming galaxies by BPT criteria. Multiwavelength data (particularly in extinction-free regimes) are needed to assess the effect of dust and the true contribution from low-luminosity or obscured AGN.

\vspace{-6pt}

\begin{figure}[H]

\includegraphics[width=.8\textwidth]{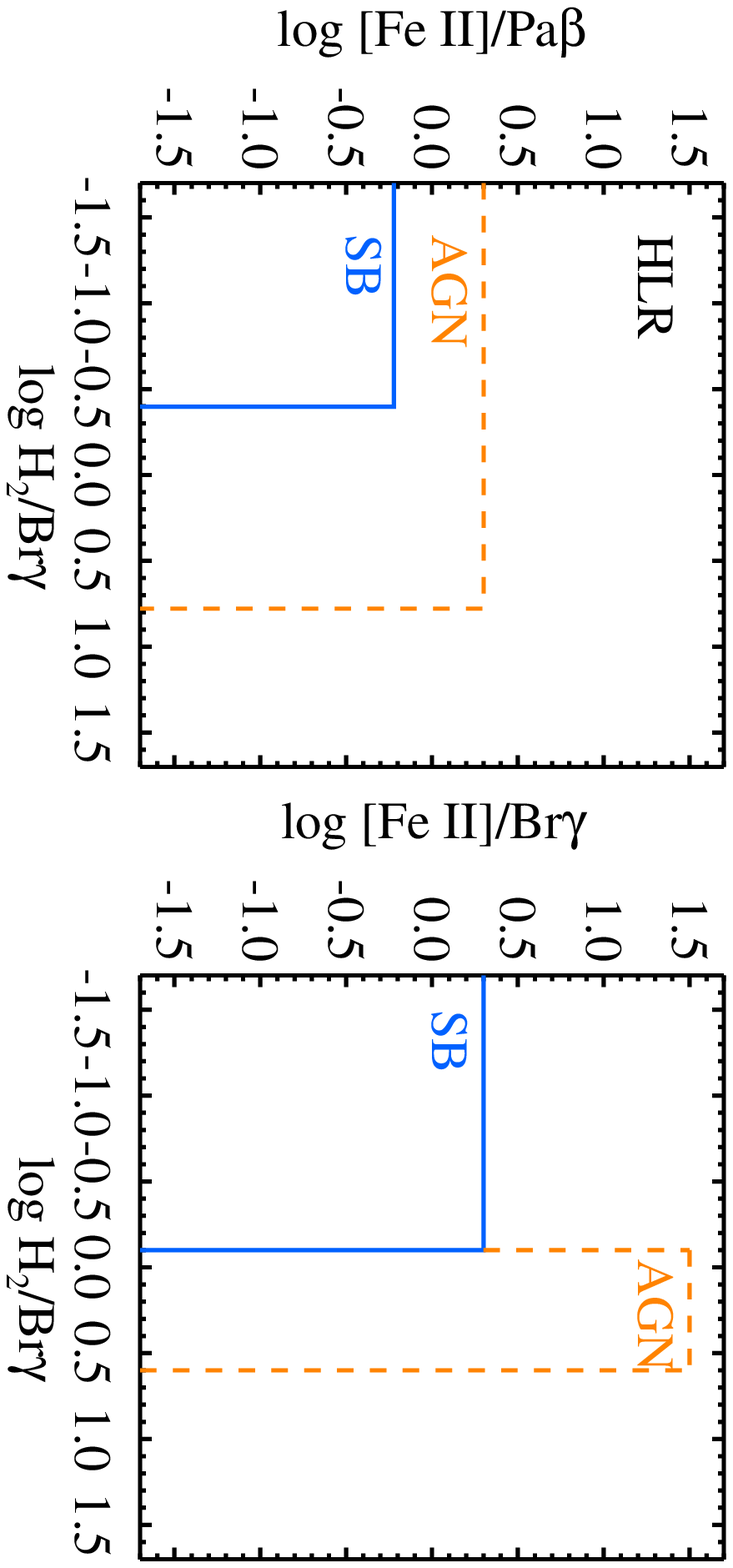}
\caption{Two near-infrared diagnostic diagrams to distinguish between excitation from starburst (SB; blue) and AGN (orange) as well as the high line ratio (HLR). The left panel shows the \feii/\pab~versus \molhy/\brg~diagram in the logarithmic scale, as derived from~\citet{Riffel13,riffel_gemini_2021}. The right panel shows the \feii/\brg~versus \molhy/\brg~diagram in the logarithmic scale, following~\citet{Colina15}. These boundaries are derived from spatially-resolved observations of nearby Seyfert and LIRGs, respectively. 
\label{fig:nir_diag}} 
\end{figure}  

The infrared wavelength range offers an advantage over the bluer regime in its ability to peer through dust, particularly important in the nuclear regions of merging galaxies. The extinction in the $K$ band measured using \brg/\brd~by~U et~al.~\cite{u_keck_2019} is $A_K \sim$ 2.5--4 mag, which corresponds to $A_V \sim$ 3--17 mag and is consistent with the $A_V \sim$ 5--40 mag found in other adaptive optics (AO) studies of LIRG nuclei~\citep{Mattila07,Vaisanen17}. 
\emph{HST} NICMOS observations of 30 (U)LIRGs reveal that the \paa~morphology is dominated by the nuclear emission with sizes of $\lesssim$1 kpc, which may be related to the AGN in some cases~\citep{Alonso-Herrero06a}. A similar, larger study of \pab~emission in 48 local (U)LIRGs observed with \emph{HST} WFC3 finds resolved nuclear clumps down to the resolution limit of 90 pc pixel$^{-1}$, with the majority having effective radii of $\sim$400 pc and consistent with being nuclear star clusters~\citep{Larson20}. 

The high-resolution resolving capability of AO-assisted integral-field unit (IFU) instruments is key to accurate measurements of the Brackett and other emission lines at the very center over globally integrated values. Similar to the optical BPT diagram, empirical emission-line studies have produced near-infrared diagnostic diagrams that assess the excitation and ionization conditions of the gas~\citep{Larkin98,rodriguez-ardila_molecular_2004,Rodriguez-Ardila05,Riffel06,Riffel13,Colina15}. Since the iron in the ISM is highly depleted onto grains, strong \feii~emission is often associated with shock-excited gas where the grains have been processed by winds, supernovae, or other ionizing sources. \mbox{Figure~\ref{fig:nir_diag}} illustrates the use of line ratios, such as \feii~(1.26 $\upmu$m)/\pab~and \molhy~1$-$0 S(1)/\brg~to differentiate among starbursts, AGN, and high-line-ratio (HLR) objects~\citep{riffel_gemini_2021}. The HLR region corresponds to where the values of both \feii/\pab~and \molhy/\brg~are high, as defined by \citet{Riffel13}. The combination of \molhy/\brg~and kinematics using resolved IFS can further identify shocked gas and outflows~\citep{u_keck_2019}. Measurements of these molecular lines will become increasingly common with the advent of \emph{JWST}; detailed theoretical models that integrate the complexities of molecular physics to radiative transfer will be needed to interpret the line-ratio results.

Near-infrared coronal lines, such as \sivi $\lambda$1.96 $\upmu$m, [Al \textsc{ix}] $\lambda$2.04 $\upmu$m, and [Ca \textsc{viii}] $\lambda$2.32 $\upmu$m~have been detected in circumnuclear regions around Seyfert nuclei~\citep{Muller-Sanchez06,Muller-Sanchez11,Piqueras-Lopez12,Bohn21}. As high-ionization lines with ionization potential $>$ 100 eV, these lines can be produced by fast shocks and/or a hard continuum, originating from a region in between the broad and narrow line regions (NLRs). They are considered robust indicators of AGN-driven outflows, though caution should be exercised in associating coronal line detection with a central point source in merger nuclei where the nuclear structure may be more complex in nature. In cases where a late-stage merger may harbor a mixture of star clusters and potentially (multiple) remnant nuclei, the mere detection of a coronal line from integrated-light spectroscopy may not correctly indicate where the outflow-driving AGN is located. This is exemplified in the ultraluminous infrared late-stage merger Mrk 273: the identified outflow based on the detection of \sivi~emission did not come from the X-ray AGN but rather appears to originate from, and helps lend evidence to the presence of, a second, obscured CT AGN~\citep{iwasawa_location_2011,u_inner_2013} whose AGN nature was only confirmed later with deep Chandra data~\citep{liu_elliptical_2019}. Despite its documented detection in several (U)LIRGs~\citep{Muller-Sanchez11, Rodriguez-Ardila11, u_inner_2013}, the \sivi~line was not detected in a single-slit spectroscopic survey of 42 GOALS LIRGs~\citep{Borish17} using TripleSpec~\citep{Herter08} 
on the ARC 3.5-m Telescope and the Palomar 200-inch Hale Telescope. A likely explanation is that the \sivi~outflow is compact (e.g., $\sim$400 pc in the case of Mrk 273~\citep{u_inner_2013}) and its signal is thus diluted in large-aperture ($\sim$1\arcsec) spectroscopy as opposed to subarcsecond resolved observations of the central regions.

\subsection{Probing the Infrared Excess}
\textls[-15]{The mid- and far-infrared are crucial for diagnosing AGN at various obscuration levels for a complete census of AGN activity in (U)LIRGs. For focused discussions of the characteristics of mid-infrared AGN properties and the observational studies of (U)LIRGs using the \emph{Spitzer Space Telescope}, we refer the readers to recent reviews by \citet{lacy_active_2020,armus_observations_2020},} and Sajina {et~al.}~\citep{Sajina22} in this Special Issue. Here, we briefly mention some \emph{Spitzer} and \emph{Herschel} results on GOALS (U)LIRGs, including the AGN fraction to the mid-infrared and total luminosity measured by~\citet{diaz-santos_herschel_2017} and AGN properties from high ionization potential lines, such as \nev~presented by~\citet{petric_mid-infrared_2011}.

Among the GOALS sample, 18\% of all LIRGs contain an AGN based on \nev~detection~\citep{petric_mid-infrared_2011}. Diagnostics using the \nev/\neii, \oiv/\neii~ line ratios, the 6.2~$\upmu$m~PAH equivalent width, and the mid-infrared continuum have been used to compute the mid-infrared and bolometric AGN strengths~\citep{diaz-santos_herschel_2017} and indicate that in 10\% of the sources, the AGN contribution to the total infrared luminosity exceeds 50\%~\citep{petric_mid-infrared_2011}. This is consistent with another study that finds the AGN bolometric contribution to the infrared luminosities for 70\% of (U)LIRGs to be generally small: $L_\mathrm{AGN bol}/L_\mathrm{IR} \leq 0.05$; only $\sim$8\% of the targeted sources have a significant contribution $L_\mathrm{AGN bol}/L_\mathrm{IR} > 0.25$~\citep{Alonso-Herrero12}. In the local universe, the AGN bolometric contribution to the infrared luminosity does in fact increase with the infrared luminosity~\citep{Veilleux95,Desai07,petric_mid-infrared_2011}, even if the impact of the AGN on the far-infrared luminosity of the mid-infrared dominated AGN LIRGs is very limited~\citep{Diaz-Santos13}. The presence of an AGN compacts the spatial extent of the mid-infrared host dust continuum but has a negligible effect on the PAH-emitting regions~\citep{Diaz-Santos11}.
IRAC color selection techniques~\cite{Stern05,Lacy04} might miss 36--50\% of the AGN-dominated (U)LIRGs or include false positives in starburst-dominated ones; 
photometric selections based on the \emph{Wide-field Infrared Survey Explorer (WISE)} colors~\citep{Satyapal18,Pfeifle21} and high ionization lines~\citep{Satyapal21} will be useful in selecting obscured or CT AGN, in particular.
These mid-infrared identification techniques will be effective with \emph{JWST}'s superb resolution and sensitivity over \emph{Spitzer}'s, where the emission from compact point sources might be diluted within the beam or nuclear sources below the confusion limit appeared blended. 

\subsection{The Submillimeter and Radio Regimes with ALMA and VLA} 
In the extinction-free submillimeter and radio wavelength ranges, excellent spatial resolution can be achieved with interferometry---down to $\sim$ 0\farcs01 (0\farcs04) 
in the high-frequency bands for ALMA (VLA). This capability allows us to bypass the dust columns and resolve the energetically dominant regions in the dusty cores of local (U)LIRGs, distinguishing between recent star formation and AGN activity. Often, star formation is taken to emit free--free, or ``thermal'' emission while AGN radiates synchrotron or ``non-thermal'' emissions. However, the non-thermal radio emission can be a tracer for star formation activity due to synchrotron resulting from particle acceleration in supernovae, 
whereas the non-thermal emission of AGN may be a mixture of synchrotron emission and that of an accretion disk, which radiate thermally. 
Complementing multiwavelength data sets, a combination of the location of the radio emission peak (though this is complicated by the interaction mechanisms, which could displace the central SMBH), morphology,  compactness, and brightness temperature is often used to help verify the AGN nature of the radio source.
The physics of ULIRGs with MUSE and ALMA~(PUMA; \cite{pereira-santaella_are_2021}) project observed 220 GHz continuum and CO(2-1) gas in 22 $z < 0.165$ ULIRGs (32 individual nuclei) to address whether ULIRGs are powered by AGN. Determining the nuclear $L_\mathrm{IR}$ and cold molecular gas surface densities to be $\Sigma_{L_\mathrm{IR}} = 10^{11.5-14.3}L_\odot$ kpc$^{-2}$ and $\Sigma_{H_2} = 10^{2.9-4.2} M_\odot$ pc$^{-2}$, respectively, the derived $\Sigma_{SFR}$ would imply extremely short depletion times and unphysical star-formation efficiencies for 70\% of the sample,  arguing in favor of the presence of obscured AGN in these nuclei. 

The claim that ULIRGs host obscured AGN is further evidenced by ALMA observations of dense molecular gas tracers HCN $J = 3-2$ and HCO$^+~J=3-2$, where the enhanced HCN emission in the nuclear region may be explained by the presence of a luminous buried AGN, high molecular gas density and temperature, and/or mechanical heating by spatially compact nuclear outflows~\citep{imanishi_alma_2019}. The link between a heightened ${L}_\mathrm{HCN(1-0)}^\prime /{L}_\mathrm{{HCO^+}(1-0)}^\prime$ ratio (HCN/HCO$^+$ hereafter) and the presence of AGN in (U)LIRGs has previously been investigated by~\citet{Kohno03,Costagliola11} and \citet{Privon15}. While AGN generally has high HCN/HCO$^+$, the reverse does not seem to be true: high HCN/HCO$^+$ values could be driven by sources other than the AGN~\citep{Privon15,Privon20}. Hard X-ray observations of four mid-infrared classified starburst galaxies with high HCN/HCO$^+$ ratios did not show that they harbor heavily obscured AGN, and for hard X-ray-identified AGN, the inferred AGN luminosity and AGN fractions were uncorrelated with HCN/HCO$^+$ ratios~\citep{Privon20}. One viable mechanism for driving the increased HCN/HCO$^+$ ratio may be mechanical heating from shocks or outflows, which could conceivably originate from an AGN---but the lack of a direct link between high HCN/HCO$^+$ sources and AGN hosts suggests that the mechanical heating may have plausibly come from starburst-driven outflows. 

High-resolution ($\leq$ 0\farcs1, or 30 pc) studies of dense gas enabled by ALMA and VLA have facilitated the search for compact obscured nuclei (CONs) in (U)LIRGs~\citep{Sakamoto13}. Using the rotational line of HCN in its vibrationally excited $v_2 = 1$ state (hereafter HCN-vib), whose first extragalactic detection was in LIRG NGC 4418 by~\citet{Sakamoto10}),~\citet{Falstad21} searched for HCN-vib in 46 (U)LIRGs and identified CONs in 38$^{18}_{13}$\% of the ULIRGs, 21$^{12}_{6}$\% of the LIRGs, and 0$^{9}_{0}$\% of the lower luminosity galaxies. The HCN-vib emission is too luminous with respect to the infrared luminosity to represent a normal cool mode of star formation~\citep{Aalto15}. Instead, the intense HCN-vib emission may come from CT AGN or from an embedded, compact, hot ($T > 200 K$), and opaque starburst~\citep{Andrews11}. Although the jury is still out on the exact nature of CONs, the evidence that CONs seem to be exclusively hosted in (U)LIRGs suggests that they may be an extreme example of obscured AGN. In a case study by LIRG IC 860, \citet{Aalto19} notes that the infrared core (within $r < 10$ pc) is consistent with a buried, efficiently accreting SMBH, or an extreme, O-star only starburst. Whether and how HCN-vib in CONs may correlate with outflow properties is relevant for understanding the nuclear feedback mechanism in the context of CONs being a heavily obscured stage of evolution~\cite{Falstad19}, a topic we will revisit in Section~\ref{sec:feedback}.

A recent study on the radio properties of nearby ULIRGs using the VLA suggested that ULIRGs have common radiative processes regardless of the presence of optical AGN~\citep{hayashi_radio_2021}. Based on high angular resolution (0\farcs05$-$0\farcs2), 
imaging of 22 (U)LIRGs at 33 GHz, nearly half of the samples were found to host single bright compact sources that contribute $>$ 50\% of the total emission~\citep{barcos-munoz_33_2017}, and nearly half of the subsamples are either confirmed AGN or potential AGN candidates, given their resolved compact sizes. The fact that the sample selection here focused on ULIRGs whose AGN were not identified by an optical signature ($\sim$70\% of all the ULIRGs) lends support to the idea that the AGN might contribute equally to every ULIRG. Multiwavelength diagnosis combining radio and far-infrared, for instance, offers further insight into the nature of the compact sources by assessing whether the infrared light powered by starbursts or AGN would correspond to free--free or synchrotron emission.


\subsection{Spectral Energy Distributions and Multiwavelength Comparisons}
Given the presence of AGN, stars, and dust at play in the complex nuclei of (U)LIRGs, cross-wavelength calibration is useful to evaluate the effectiveness and complementarity of individual AGN-selection approaches and to achieve a complete census of active nuclei within the (U)LIRG population. Spectral energy distribution (SED) studies of nearby (U)LIRGs provide the means to investigate the multiband emission from various galactic subcomponents as the galaxies progress through the merger sequence~\citep{u_spectral_2012,Paspaliaris21}. With a sample of 65 (U)LIRGs from GOALS, \citet{u_spectral_2012} amassed the aperture-matched X-ray to radio-integrated photometry and compared several of the multiwavelength indicators for~AGN: 
\begin{itemize}
    \item The correlation between far-infrared and radio emissions as a way to distinguish radio-loud AGN from starbursts~\citep{Condon91,Ji00,Donley05}; while radio excess (infrared-excess) sources with extremely low (high) values of $q$ (as defined in Table~\ref{tbl:multiwavelength}) could both contain an AGN or obscured AGN, there are more likely to be starburst contaminants among the high $q$ sources~\citep{Yun01,Vardoulaki15}. This indicator has since been refined by adopting 24 $\upmu$m~flux instead of far-infrared flux~\citep{Appleton04,Donley05}.
    \item The X-ray luminosity threshold or hardness ratio based on relative contributions from the hard (2--8 keV) and soft (0.5--2 keV) bands~\citep{Iwasawa93,Mushotzky93}; the X-ray---radio excess criterion due to AGN activity~\citep{Alberts20};
    \item Mid-infrared power-law index and \emph{Spitzer}/IRAC colors (see Sajina {et~al.} 2022~\citep{Sajina22} and Lyu and Rieke 2022~\citep{Lyu22} in this Special Issue for more in-depth discussions); and 
    \item Optical spectral classification based on the BPT diagrams and line widths of the hydrogen recombination emission features~\citep{Yuan10}
\end{itemize}

\begin{figure}[H]
\includegraphics[width=0.7\textwidth]{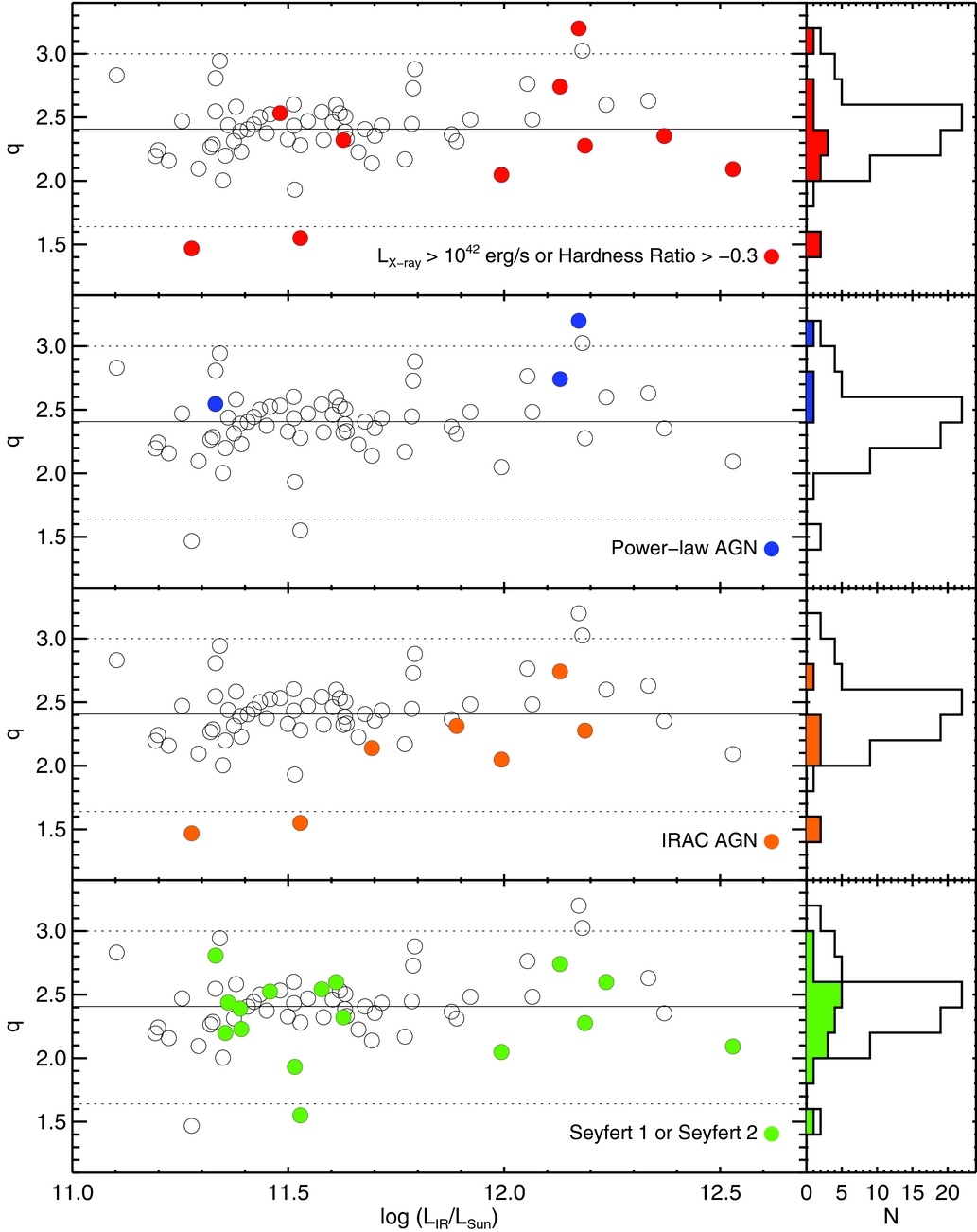}
\caption{Radio-infrared flux ratio $q$ vs. $\nu L_\mathrm{IR}$ plots are shown for a sample of local (U)LIRGs along with the distribution in $q$ on the right panels. The solid line indicates the median value ($q$ = 2.41) of the sample, with $\sigma$ = 0.29. The dotted lines indicate the radio excess ($q$ $<$ 1.64) and infrared-excess \mbox{($q$ $>$ 3.0)} criteria from~\citet{Yun01}. From top to bottom are four different AGN indicators (filled circles/histograms): (red) X-ray luminosity and hardness ratio, (blue) mid-infrared power-law slope, (orange) IRAC colors, and (green) optical spectral classification. Figure from~\citet{u_spectral_2012}; reproduced with permission from the author.
\label{fig_u12}}  
\end{figure}

The multiwavelength indicators along with the associated AGN criteria are summarized in Table~\ref{tbl:multiwavelength}. 
The results from the cross-wavelength comparison of these various selection techniques show that they complement each other (see also Lyu {et~al.} in the review). Approximately 60\% of the ULIRGs and 25\% of the LIRGs were classified to host at least one AGN by at least one of the aforementioned criteria (Figure~\ref{fig_u12};~\citep{u_spectral_2012}). 

While a luminous AGN, if present at the center of a galaxy, tends to dominate the light of the host galaxy if its luminosity is above $10^{44}$ erg s$^{-1}$~\citep{prieto_spectral_2010}, (U)LIRGs may harbor lower-luminosity or dust-obscured AGN whose emissions may contribute only negligibly to global SEDs.  A more definitive approach to understanding the growth pathway of a SMBH through the merging sequence would be a resolved SED study that isolates the nuclear emission with an aperture size that minimizes contamination from the host galaxy (Gao, U {et~al.}, in preparation). An even more powerful capability is the NIRSpec IFU, where the presence of hot dust can be isolated spatially and through the dilution of the stellar absorption lines. The unprecedented high-resolution imaging and spectroscopic capability of \emph{JWST} in the near- and mid-infrared will be valuable in constraining the AGN hot dust emission for such studies.

\begin{table}[H] 

\caption{A summary table for multiwavelength AGN diagnostics in (U)LIRGs\label{tbl:multiwavelength}}
\setlength{\cellWidtha}{\fulllength/3-2\tabcolsep+0.6in}
\setlength{\cellWidthb}{\fulllength/3-2\tabcolsep+0.6in}
\setlength{\cellWidthc}{\fulllength/3-2\tabcolsep-1.2in}

\begin{adjustwidth}{-\extralength}{0cm}
\centering 
\scalebox{1}[1]{\begin{tabularx}{\fulllength}{>{\centering\arraybackslash}m{\cellWidtha}>{\centering\arraybackslash}m{\cellWidthb}>{\centering\arraybackslash}m{\cellWidthc}}
\toprule
\textbf{Multiwavelength Indicator}	& \textbf{AGN Criteria}	& \textbf{Ref.}\\
\midrule
\multicolumn{1}{l}{Radio-infrared flux ratio} & & \\
$q \equiv \log\left(\frac{\mathrm{FIR}}{3.75\,\times\,10^{12}~\mathrm{W m}^{-2}}\right) - \log\left(\frac{S_\mathrm{1.4~GHz}}{\mathrm{W m}^{-2} \mathrm{Hz}^{-1}}\right)$ &
$q < 1.64$: radio-excess & \citep{Condon91,Yun01,Vardoulaki15} \\
FIR $\equiv 1.26 \times 10^{-14} (2.58~S_{60\upmu \text{m}}+S_{100\upmu \text{m}}) \mathrm{W m}^{-2}$  & & \\
or & & \\
$q \equiv \log(f_{24\upmu \text{m}} / f_\mathrm{1.4 GHz})$ & $q < 0$ & \citep{Appleton04,Donley05} \\
\midrule
\multicolumn{1}{l}{X-ray Detection} & & \citep{Kartaltepe10} \\
 & $L_\mathrm{2-10~keV} > 10^{42}$ erg s$^{-1}$  & \\
\multicolumn{1}{l}{X-ray Hardness Ratio} & & \citep{iwasawa_c-goals_2011} \\
\begin{tabular}{@{}c@{}}HR $= (H-S)/(H+S)$ \\ $H$ ($S$): counts in 2$-$8 (0.5$-$2) keV\end{tabular} & HR $> -0.3$ & \\
\multicolumn{1}{l}{X-ray Radio Excess} & & \citep{Alberts20} \\
 & $L_\mathrm{0.5-7~keV}/L_\mathrm{6 GHz} > 1.2 \times 10^{19}$  & \\
\midrule
\multicolumn{1}{l}{Mid-infrared Power-law Index}	& 	& \citep{Kartaltepe10,Alonso-Herrero06b,Donley07} \\
 & $\log(\nu L_{4.5}) - \log(\nu L_{2.2}) > 0$	 & \\
\midrule
\multicolumn{1}{l}{IRAC Colors}		& 			&  \citep{Donley12} \\
$x = \log \left(\frac{f_{5.8\upmu \text{m}}}{f_{3.6\upmu \text{m}}}\right)$, $y = \log \left(\frac{f_{8.0\upmu \text{m}}}{f_{4.5\upmu \text{m}}}\right)$ & 
\begin{tabular}{@{}c@{}c@{}}
$x \geqslant 0.08 \wedge y \geqslant 0.15 $ \\
$\wedge y \geqslant (1.21 \times x) - 0.27 \wedge y \leqslant (1.21 \times x) + 0.27 $ \\
$\wedge f_{4.5\upmu \text{m}} > f_{3.6\upmu \text{m}} \wedge f_{5.8\upmu \text{m}} > f_{4.5\upmu \text{m}} \wedge f_{8.0\upmu \text{m}} > f_{5.8\upmu \text{m}}$ \end{tabular} & \\
\midrule
\multicolumn{1}{l}{Optical Spectral Classification} &  & \citep{Yuan10,Kewley01} \\
 & Seyfert 1: \ha~(FWHM) $> 5 \times 10^3$ km s$^{-1}$ & \\ 
 & Seyfert 2: lie above \citet{Kewley01} theoretical lines in BPT diagrams &  \\ 
\bottomrule
\end{tabularx}}
\end{adjustwidth}
\end{table}

\section{The Coevolution of SMBHs and the Host Galaxies}
\label{sec:feedback}

\begin{figure}[H]

\includegraphics[width=.95\textwidth]{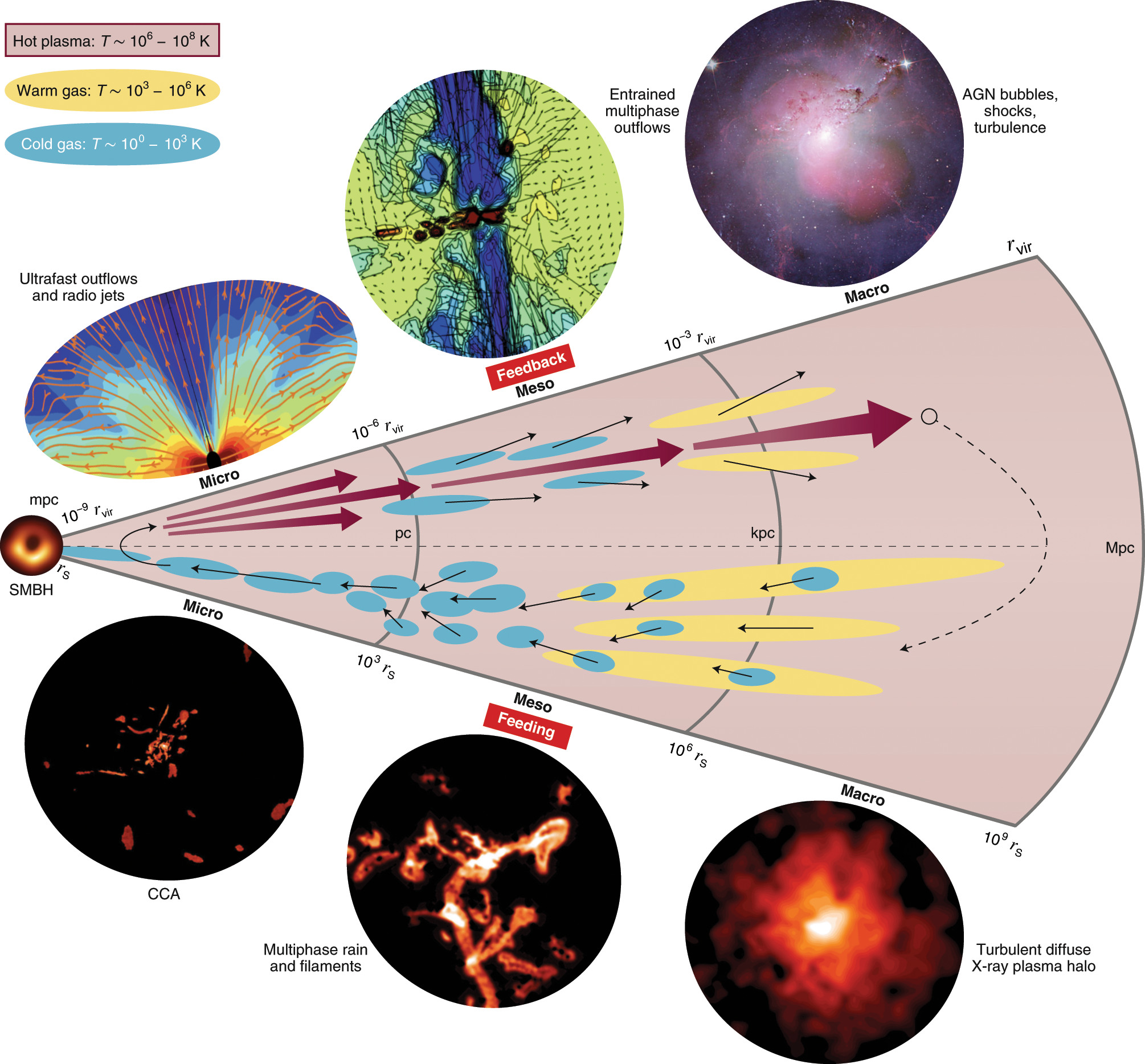}
\caption{Schematic showing the multiphase, multiscale fueling, and feedback processes in a self-regulated cycle. This diagram captures several major crucial phases from SMBH scales (sub-pc) to the virial radius of a galaxy (Mpc). The arrows illustrate the anticipated motions of the multiphase gas with different temperatures (red: hot plasma at $T \sim 10^6-10^8$K; yellow: warm gas at $T \sim 10^3 - 10^6$ K; blue: cold gas at $T \sim 10^0 - 10^3$ K). This review features results from observational studies on gas accretion onto SMBH and entrained, multiphase outflows from the nuclear region to galactic scales in nearby (U)LIRGs. Figure from \citet{Gaspari20}; reprinted with permission.
\label{fig:feeding_feedback}} 
\end{figure}

The theoretical and observational studies of black hole feeding and feedback have provided key insights toward understanding galaxy evolution as part of the cosmic ecosystem. These fueling and feedback processes are responsible for many of the correlations between the observed black hole mass, SFR, and other galaxy properties seen in statistical studies of SMBHs and their host galaxies. Building on the assumption that SMBHs and their hosts are linked, how gas of different phases and temperatures may be shocked, compressed, or entrained throughout the ISM at different scales comes into the spotlight (Figure~\ref{fig:feeding_feedback}; see also \citet{Gaspari17_unifying}). Here, we review recent observational results from integral-field spectrographs and interferometry arrays that focus on nearby (U)LIRGs, many of which prominently feature these energetic events to be examined in detail. 


\subsection{AGN Fueling}

\subsubsection{Transporting Gas to the Center}

SMBHs feed on gas that may have originated from the galactic halo at 50 kpc scales via cold accretion flows~\citep{Gaspari17_raining}. Simulations of merging gas-rich disk galaxies have shown that major mergers are generally more effective in transporting gas to the nuclei than their higher-mass ratio merger counterparts, resulting in 50\% as opposed to 20--30\% of the gas being funneled to the center of the system~\citep{Naab01}. More recent galaxy merger simulation suites using hydrodynamic simulations coupled with dust radiative transfer have uncovered obscured signatures of merger-triggered AGN fueling~\citep{Blecha18}. The obscuration phase coincides with the peak of BH fueling, such that merger-triggered AGN fueling could be missed among late-stage mergers. While mid-infrared AGN selection might include contribution from star formation, gas that fuels starbursts also fuels AGN simultaneously in merger nuclei. Using the Stars and MUltiphase Gas in GaLaxiEs (SMUGGLE~\citep{Marinacci19}) model that incorporates the multiphase ISM and stellar evolution, \citet{Sivasankaran22} demonstrated that interaction-induced inflows cause peak gas densities near the SMBH to increase by orders of magnitude, subsequently resulting in elevated and bursty star formation as well as enhanced accretion. The instantaneous accretion rate depends strongly on the conditions of the local ISM, the constraints for which will be derived from high-resolution observations of galactic nuclei. 

Specifically, in nearby infrared galaxies, large reservoirs of cold and warm molecular gases have been detected in late-stage interacting galaxies as amassed by the merging mechanism. Submillimeter interferometry studies of merger remnants have found large amounts of CO gases (10$^{7-11} M_\odot$) that reside in rotation-supported turbulent disks varying from 1.1 to 9.3 kpc in size \citep{Ueda14}. Within the central kpc regions, young stellar and warm \molhy~disks are nearly ubiquitous among ULIRGs. With stellar ages $<$ 30 Myr, masses of 10$^{8-10} M_\odot$, and effective radii $r_\mathrm{eff} \approx$ 200--1800 pc~\citep{medling_stellar_2014}, these circumnuclear molecular gas disks have also been seen in simulations of galaxy mergers hosting inflow rates as high as 10$^4 M_\odot$ year$^{-1}$ \citep{Mayer10,mayer_massive_2013}. The presence of this gas signals sufficient fueling for forming stars in situ during the merger, and this gas would likely be driven toward the center to aid the final coalescence of binary black holes in short timescales ($<$$10^7 M_\odot$) \citep{Escala05,Dotti07,Mayer07}. 

\subsubsection{Black Hole Masses in Galaxy Mergers}

\begin{figure}[H]

\includegraphics[width=.98\textwidth]{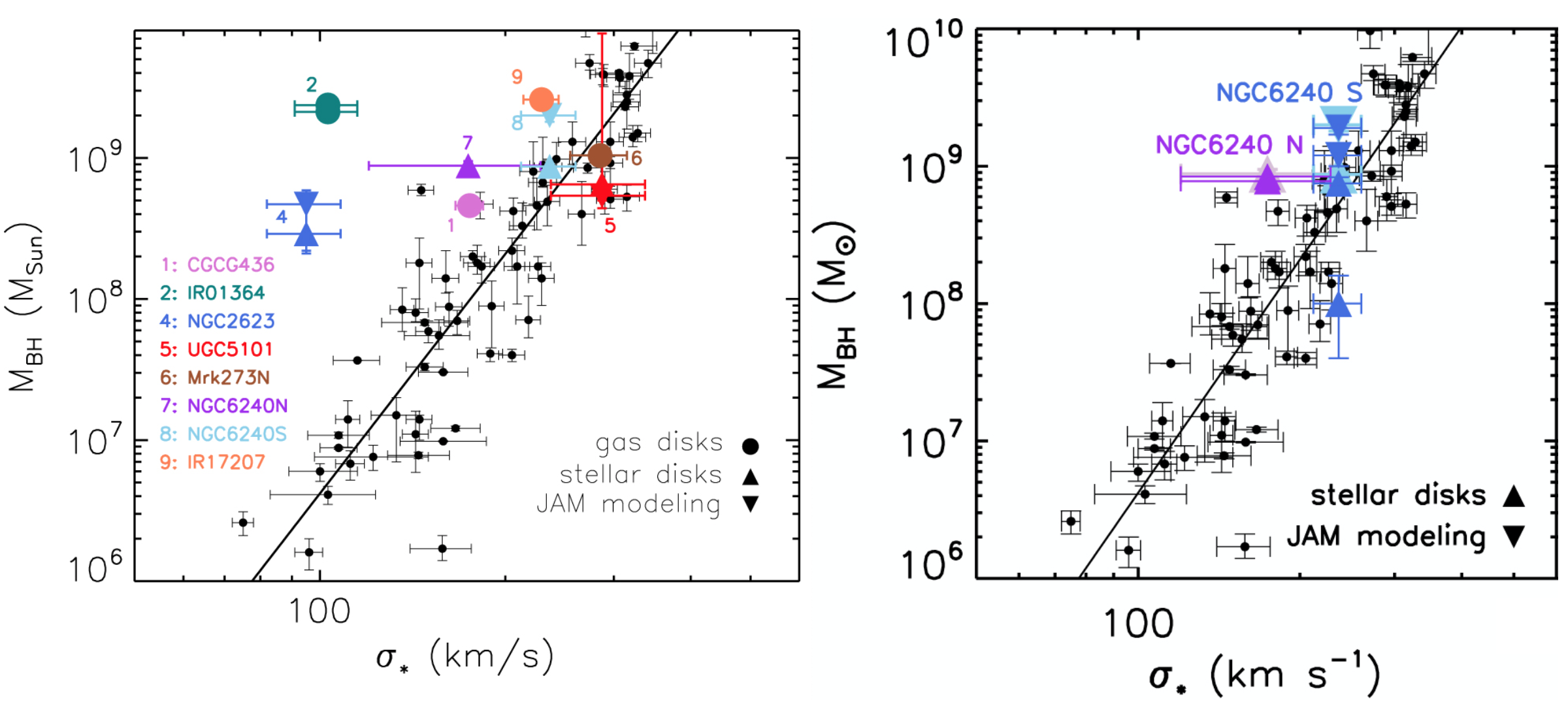}
\caption{(Left) The $M_\mathrm{BH}-\sigma$ relation for normal galaxies from \citet{McConnell13} (black) and merging (U)LIRGs (colored) from different mass determination methods. (U)LIRGs appear to lie on top of or above the scaling relation. Figure from \citet{medling_following_2015}. (Right) The same plot but now the black hole mass measurement for NGC 6240 has been updated to incorporate correction based on the submillimeter measurement of the gas mass. Figure adapted from \citet{medling_how_2019}. Both panels~\textcopyright~AAS, reproduced with permission.
\label{fig:msigma}} 
\end{figure}

The presence of nuclear disks provides a means to determine dynamical black hole masses where the sphere of influence is resolved. Local galaxies ($<$20 Mpc) are prime candidates for these dynamical measurements, given the resolving capability of the current integral-field spectrographs and interferometric facilities. The nearby (U)LIRGs are slightly further, but they are also more massive (with mean stellar mass $\log (M_\star/M_\odot) = 10.79 \pm 0.040$; \citep{u_spectral_2012}) with expectedly larger black holes with larger spheres of influence, so dynamical measurements are still plausible up to $z \lesssim 0.05$, or 200 Mpc. 
Modeling the high-resolution integral-field data that demonstrate the primarily Keplerian rotation of the central point source in molecular and ionized gas, \citet{medling_following_2015} suggest that late-stage merging (U)LIRGs tend to harbor overmassive black holes relative to the black hole--galaxy relationships based on normal galaxies: $M_\mathrm{BH}-\sigma$ (see the left panel in Figure~\ref{fig:msigma}), $M_\mathrm{BH}-L_\mathrm{bulge}$, and $M_\mathrm{BH}-M_\star$ \citep{Ferrarese00,Gebhardt00,Tremaine02,Merritt01,Kormendy01,marconi_relation_2003,Magorrian98,Gultekin09,McConnell13}. 

Whether SMBH growth truly precedes that of the bulge remains nuanced (c.f.~\citep{Kormendy13}), where the observed relationship between the two might depend on the sample selection (e.g., isolated vs. mergers), host properties (e.g., dry mergers), or measurement uncertainties (e.g., systematics in measuring black hole masses and/or bulge velocity dispersion). In fact, one caveat with reporting the stellar velocity dispersion of the bulge in galaxy mergers is that it is difficult to define what is considered the bulge when the progenitors are intertwined. Simulations have also shown that the velocity dispersion oscillates during the course of the merger~\citep{Stickley14}, easily introducing uncertainties to the scaling relation. The accuracy of the black hole mass measurement is also susceptible to systematics associated with the various mass determination methods. 

Further investigating the phenomenon of overmassive black holes, whose masses are postulated to include excess gas mass within the sphere of influence that has yet to be accreted onto the central source, \citet{medling_how_2019} observe one such merging system---NGC 6240---with high-resolution ALMA Band 6 observations that probe at the physical scale of 15$-$30 pc. This study computes gas mass via two approaches: 1. using the gas mass calculation technique based on the dust continuum from \citet{scoville_alma_2015}, and 2. using a resolved-calibrated $\alpha_\mathrm{CO}$ measurement. They found cold CO molecular gas reservoirs of a few times 10$^7$ and 10$^8 M_\odot$ for the north and south nuclei, respectively. These measurements provide a black hole mass correction that potentially brings the dynamical measurement based on near-infrared gas and star tracers closer to the M-$\sigma$ relation (see the right panel in Figure~\ref{fig:msigma}). If the detected CO gas mass has yet to be accreted onto the central SMBH, it is thus posited that the black hole may not, after all, be overmassive relative to the property of the bulge. Rather, we can deduce the timescale by which the gas is accreted given finer time sampling of black hole mass measurements along the merger sequence. While evidence for merger-induced black hole fueling is present, SMBH studies with larger samples of mergers will be needed to resolve the scaling relations conundrum. 

\subsubsection{Emergence of AGN in ULIRGs} 
The GOALS sample encompasses (U)LIRGs with $\log L_\mathrm{IR} = 11.0 - 12.5$; among which the brightest object is Mrk 231, the only bonafide QSO in the sample~\citep{armus_goals_2009}. Because of their relatively moderate luminosities compared to those of quasars, plus the mixed presence of starbursts and AGN that sometimes obfuscates the dominant energy driver, LIRGs are often overlooked in the context of luminous AGN studies. GOALS was designed to probe primarily the merger phases that bridge the normal star-forming galaxies to the extreme QSO populations.  Section~\ref{sec:agn} presents the detection rate of AGN and obscured AGN activity in (U)LIRGs amidst a discussion of multiwavelength identification techniques, but the AGN dominates the host galaxy light, mostly after the dust blow-out phase in the merger.  

At the higher luminosities, the average AGN contribution in ULIRGs ranges from $\sim$15\%--35\% among optically-classified H \textsc{ii}-like and LINER ULIRGs to $\sim$$75\%$ among Seyfert 1 ULIRGs, reaching $\sim$$80\%$ in QSOs~\citep{Veilleux09}. Such a comparison benefits from well-matched samples, such as those in the Quasar-ULIRG Evolution STudy (QUEST;~\citep{Schweitzer06,Veilleux08,Veilleux12}) survey, which consists of ULIRGs from the~\emph{IRAS} 1-Jy Survey of ULIRGs~\citep{Kim98} and QSOs from the Palomar--Green (PG) quasar survey~\citep{Schmidt83}. A similar trend in increasing AGN fraction with infrared luminosity was noted by~\citet{Alonso-Herrero12}, with a jump from $\sim$$50\%$ to $\sim$$80\%$ occurring at $\log L_\mathrm{IR} = 12.5$. 
A recent simulation of a galaxy merger demonstrates that the black hole accretion rate peaks correlate with pericenter encounters, such that the second and third pericenter passes cause an enhancement in the Eddington limit~\citep{prieto_black_2021}. The result of AGN fueling is that the high mass accretion rate onto compact objects will trigger AGN activity.
Further evidence of AGN prominence in ULIRGs lies in high-velocity outflows detected in sources with large AGN luminosities~\citep{Sturm11}, the effect of which on the host galaxies will be examined in the next section.

\subsection{AGN Feedback}

Energetic feedback from stars, supernovae, and accreting supermassive black holes regulates the baryonic content of a galaxy by injecting energy and momentum into its ISM and driving powerful winds from the nucleus to galactic scales~\cite{Heckman95,Veilleux05}. In the canonical picture of galaxy evolution, these winds play a key role in the chemical evolution of galaxies through metal enrichment and redistribution, and the suppression of star formation throughout the galaxy and out into the circumgalactic medium (CGM). 
AGN feedback in particular has been invoked to explain the discrepancy between the galaxy luminosity function predicted by $\Lambda$CDM and that observed at the high-luminosity end. 
Readers are referred to articles by \citet{Veilleux05,Fabian12,Rupke18,Veilleux20} for in-depth discussions of the physics of feedback and associated investigations in quasars, Seyferts, and star-forming galaxies. In this review, we focus on key questions and results from studies of feedback in (U)LIRGs and mergers. 

\subsubsection{Feedback: A Multiphased Phenomenon}
As powerful sources hosting both starbursts and AGN, (U)LIRGs (as a population) present an intriguing albeit challenging environment to disentangle the feedback effect on the ISM. 
Theoretically, the ``Feedback In Realistic Environments-2'' (FIRE-2) model~\citep{Hopkins18} resolves star formation, feedback processes, and the multiphase structure of the ISM in an extensive suite of parsec-scale galaxy merger simulations. This framework subsequently enables the study of interaction-induced ISM conditions and their dependence on orbital geometry as well as the SFR-enhancement or SFR-suppression in different local environments---underscoring the need for large, spatially-resolved galaxy merger surveys~\citep{moreno_spatially_2020}. 

\begin{figure}[H]

\includegraphics[width=.98\textwidth]{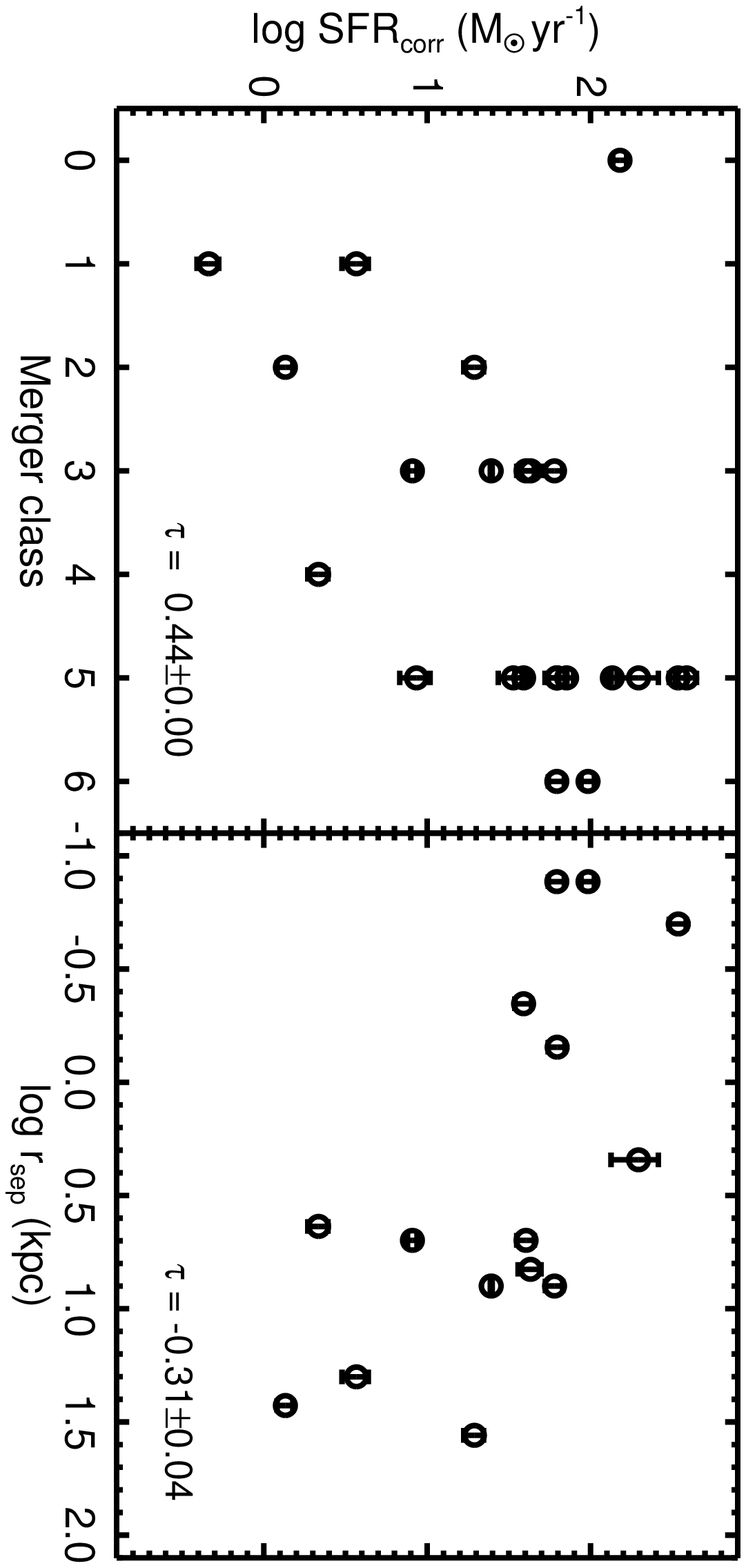}
\caption{Dust-corrected nuclear SFR on logarithmic scales as a function of the merger class (\textbf{left}) and nuclear separation $\log r_\mathrm{sep}$ (\textbf{right}). The correlation coefficient $\tau$ is given in each panel; data points reproduced here are taken from \citet{u_keck_2019} with authors' permission.
The merger classification scheme was adopted from~\citet{haan_nuclear_2011} and \mbox{\citet{kim_hubble_2013}} as follows: (0) a single galaxy with no obvious companion, (1) separate galaxies with symmetric disks and no tidal tails, (2) distinguishable progenitor galaxies with asymmetric disks and/or tidal tails, (3) two distinct nuclei sharing a common envelope, (4) closely-separated nuclei with visible tidal tails, (5) a single nucleus with prominent tails, and (6) a post-merger remnant.
\label{fig:koala_feedback}} 
\end{figure}

Observationally, massive outflows in nearby ULIRGs, often associated with late-stage mergers, appear to be ubiquitous~\citep{yamada_comprehensive_2021,pereira-santaella_are_2021}.
These transformative processes 
have been detected in ionized and neutral atomic gas~\citep{Soto12,Rupke13,Leung21}, warm molecular gas~\citep{u_inner_2013,medling_shocked_2015,u_keck_2019}, and cold molecular gas~\citep{cicone_massive_2014,Alatalo15,GarciaBurillo15,izumi_alma_2020}. 
Relative to AGN studies where outflows that correlate with decreased SFR are taken as evidence for negative feedback, the topic is much more nuanced among (U)LIRGs given the presence of both intense star formation and AGN in the nuclei, dust obscuration masking 
the intrinsic state of the AGN, and merger interaction that alters the ISM conditions on dynamic timescales. With the ability to resolve stellar and gaseous emissions on circumnuclear scales, \citet{u_keck_2019} determined dust-corrected nuclear SFRs for 21 (U)LIRGs in the GOALS sample and found that nuclear star formation exhibits a trend with the merging stage and diminishing projected nuclear separation (Figure~\ref{fig:koala_feedback}). The trends support the notion of positive feedback at small scales, consistent with the picture of merger-induced starbursts in the center of galaxy mergers~\citep{u_keck_2019}.

\textls[-15]{On top of the complexities associated with multiscale investigations of outflows, the usual challenges of constraining AGN feedback remain: determining the energetics, mass, metallicity, and physical extent of the outflowing gas will rely on resolved IFU studies~\citep{kakkad_spatially_2018,kakkad_super_2020}. }
Sources for errors in calculating mass loss rates include uncertainties in the outflow geometry and electron density in the absence of high-sensitivity and spatially-resolved data~\citep{cicone_largely_2018}. The outflowing gas can experience phase changes in different ISM environments and may be intrinsically multiphase, sharing similar kinematics but different mass, energy, and momentum contributions~\citep{pereira-santaella_spatially_2018}.
With the advent of IFUs and interferometry among both ground- and space-based facilities, proper calibrations between integrated and resolved measurements will be useful. Here we discuss a few in-depth case studies of feedback in (U)LIRGs as well as conclusions drawn from emerging surveys.


\subsubsection{A Case Study: Tracing Outflows in Mrk 273 from the Core to the Halo}

\begin{figure}[H]

\includegraphics[width=.8\textwidth]{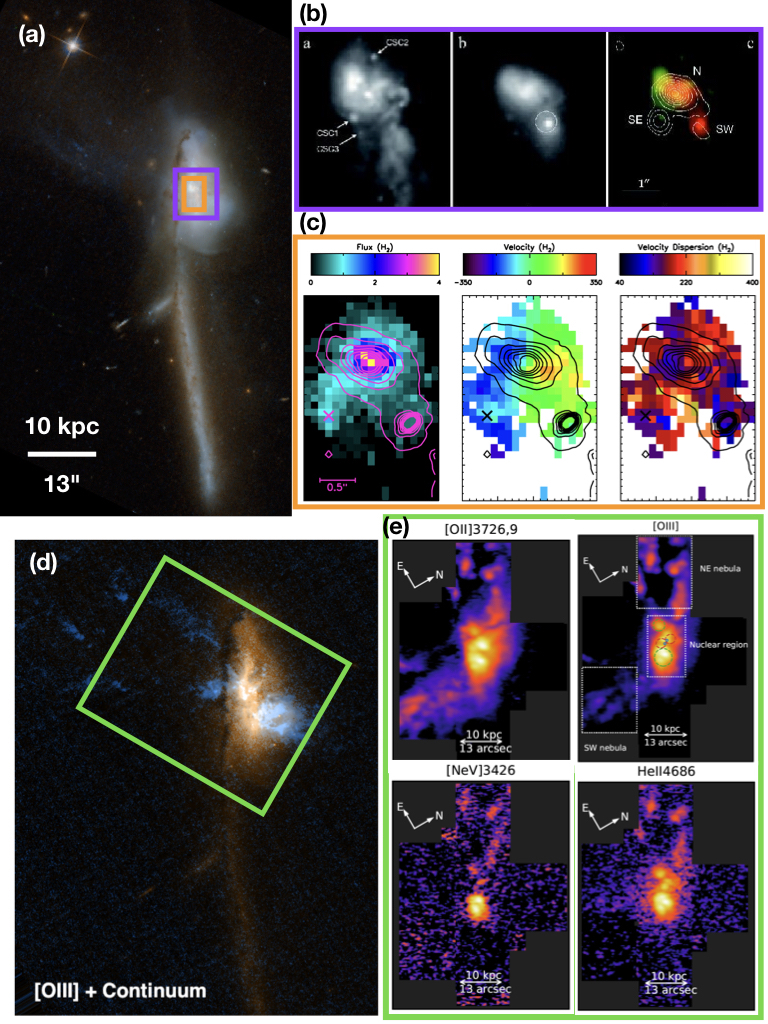}
\caption{(\textbf{a}) Mrk 273 as seen in visible light from \emph{HST}/ACS \emph{F435W} and \emph{F814W} imaging; north points up and a 10 kpc ($\sim$13\arcsec) scale bar is shown. Purple and orange boxes correspond to insets (\textbf{b},\textbf{c}), respectively; (\textbf{b}) from left to right: \textbf{a)} \emph{HST}/ACS \emph{F814W}, \textbf{b)} \emph{HST}/NICMOS \emph{F160W}, and \textbf{c)} composite of the previous two overlaid with the VLA 8.4 GHz image~\citep{Condon91} in contours, figure adapted from \citet{iwasawa_location_2011}, reproduced with permission from \textcopyright~ESO; (\textbf{c}) from left to right: flux, velocity, and velocity dispersion of the \molhy~1$-$0 S(1) overlaid with the $K$ band continuum as observed with Keck OSIRIS, figure from \citet{u_inner_2013}; (\textbf{d}) \emph{HST}/ACS continuum (red) + \oiii~(blue) \mbox{from \citet{Rodriguez-Zaurin14}}, reproduced with permission from \textcopyright~ESO; green box indicates the zoomed-in region for (\textbf{e}); (\textbf{e})~emission line maps highlighting the extent of ionized gasses (\oii, \oiii, \nev, and \heii) well beyond the nuclear region in Mrk 273; figure adapted from \mbox{\citet{Leung21}};~\textcopyright~AAS, reproduced with permission.
\label{fig:mrk273}} 
\end{figure}

Among the most nearby ULIRGs, Mrk 273 is a late-stage merger that features multiphase high-velocity AGN-driven outflows with particularly complex kinematics within its nuclear region (Figure~\ref{fig:mrk273}). Based on infrared observations from \emph{HST} NICMOS, the nuclear complex appears clumpy harboring at least one SMBH~\citep{Scoville00}. High-resolution radio observations resolved multiple sources that were identified to be nuclear star cluster candidates. \emph{Chandra} X-ray observations of 44-ks depth revealed an X-ray AGN in the southern nucleus and offered constraints that suggested the northern nucleus to be CT~(see Figure 1 in \cite{iwasawa_location_2011}). The dual-AGN nature of Mrk 273 was not confirmed until \citep{u_inner_2013} analyzed high-resolution AO near-infrared observations of the nuclear gas kinematics. In the central kiloparsec region, the warm shocked molecular hydrogen gas is found in biconical outflows emanating from the northern nucleus, where a black hole of mass $M_\mathrm{BH} = 1 \times 10^9 M_\odot$ has been located using the gas dynamical measurement~\citep{u_inner_2013}. This measurement agrees very well with that from OH maser~\citep{Klockner04}, a golden standard among black hole mass measurement techniques. The inferred density from this massive compact object exceeded that of most known stellar clusters, arguing in favor of this being a remnant nucleus from the progenitor merging galaxy. The dual AGN nature of Mrk 273 was further verified by deep 200-ks \emph{Chandra} data subsequently obtained where the hard X-ray emission was successfully resolved~\citep{liu_elliptical_2019}. There are even speculations that a gas clump shock-heated by the outflow in the \sivi~emission might actually be the third nucleus in this system~\citep{Rodriguez-Zaurin14}. The implication brought by this interesting system will be discussed in the broader context of dual SMBHs in Section~\ref{sec:duals}.

On the galactic scale, the large nebular bubble 20 kpc to the northeast direction of the nuclear region in Mrk 273 has been imaged in ionized gas via \emph{HST} narrow band imaging~\citep{Rodriguez-Zaurin14}, and more recently, ground-based IFU observations~\citep{Leung21} (Figure~\ref{fig:mrk273}). The mixing of AGN photoionization and shock excitation is detected in both the nuclear region and the NE nebula, but the relative contribution from each is assessed to be different, spatially but connected, suggesting a common origin for both. The advent of wide-field IFU observations facilitates not only kinematic studies but also resolved studies of key diagnostic line ratios. Line ratio diagnostics coupled with modeling can be a powerful technique to disentangle the excitation conditions and the intricate driving mechanisms often seen in galaxy mergers~(for example, of another in-depth study of LIRG NGC 6240 \cite{Medling21}). 

\subsubsection{How Does AGN drive outflows in (U)LIRGs?}

\begin{figure}[H]

\includegraphics[width=.7\textwidth,trim=0 0 0 300,clip]{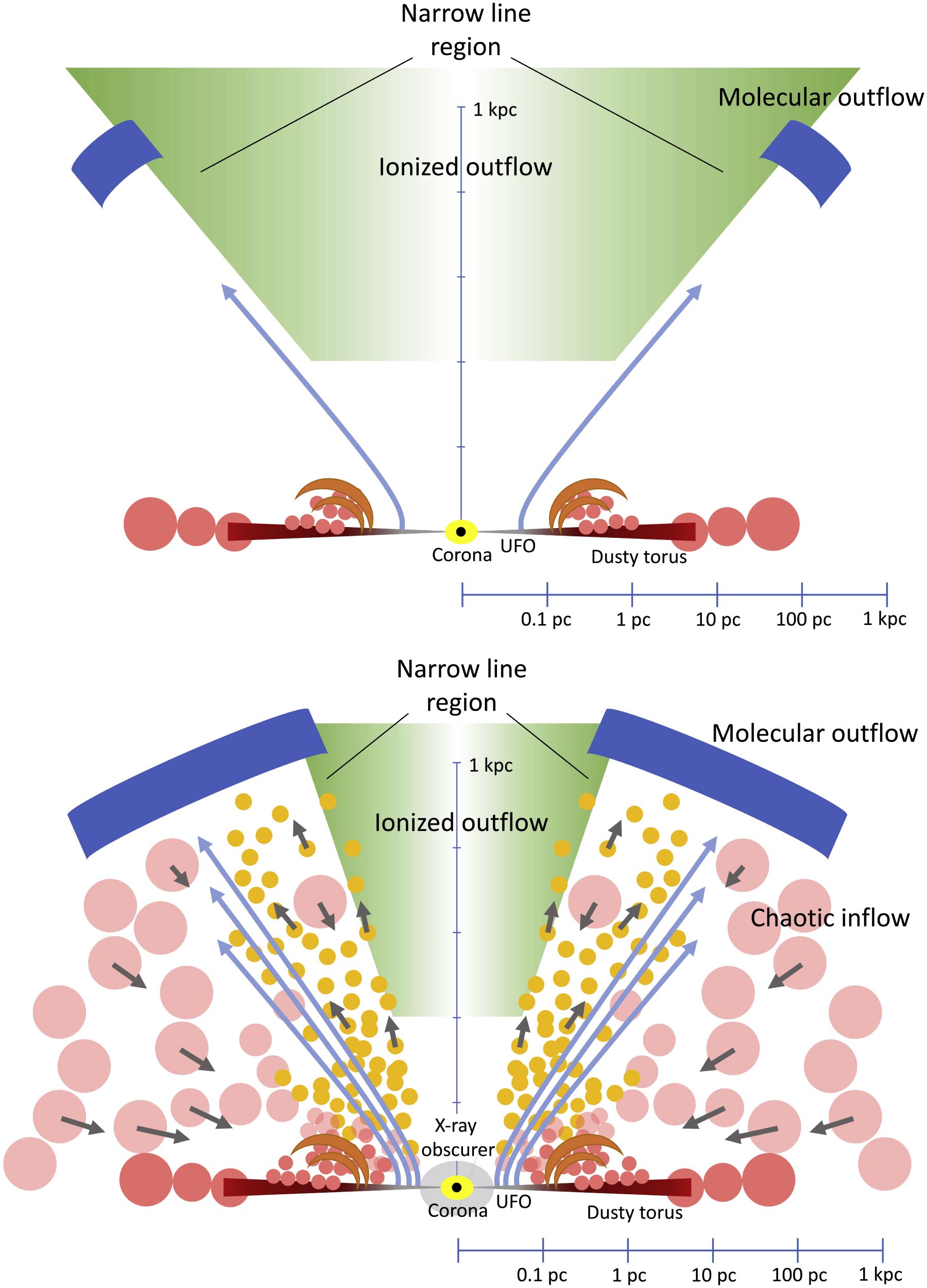}
\caption{Schematic of AGN in the late-stage merger explaining how the AGN may end up deeply buried. Galaxy interaction induces chaotic quasi-spherical inflows (light pink) toward the dusty torus (dark pink), which block the X-ray emission from the corona, and in turn, launch the UV-line driven winds (blue arrows) and drive outflowing materials (orange). These inflows and outflows are responsible for keeping the AGN obscured within molecular outflows found at larger radii.  Figure adapted from \citet{yamada_comprehensive_2021};~\textcopyright~AAS, reproduced with permission. 
\label{fig:yamada}} 
\end{figure}

 Not only are outflows multiphased and multiscaled---often requiring observations from different facilities to fully sample the processes involved---but the dynamics of mergers plus the sometimes obscured state of the AGN add challenges to putting together the pieces of the feedback puzzle. Fortunately, the picture is becoming clearer with the advent of recent multiwavelength studies being devoted to investigating the detailed mechanisms for this phenomenon in nearby (U)LIRGs. A \emph{NuSTAR} and {Swift/BAT} study found that the most luminous AGN (with $L_\mathrm{bol,AGN} \sim 10^{46}$ erg s$^{-1}$) have relatively small column densities ($\log N_H \leq 23\,\mathrm{cm}^{-2}$) 
 \citep{yamada_comprehensive_2021}. The proposed scenario depicts strong multiphase outflows triggered by chaotic quasi-spherical inflows but the AGN in these late-stage mergers might appear X-ray weak and CT due to obscuration by inflowing or outflowing material (Figure~\ref{fig:yamada}). This finding is consistent with that from a far-infrared \emph{Herschel}-based study of OH, where the highest molecular outflow velocities are found in buried sources~\citep{gonzalez-alfonso_molecular_2017}.  
 Using high-resolution ALMA images, the PUMA survey found that two-thirds of the surveyed ULIRGs have nuclear radiation pressures above the Eddington limit, giving weight to the role that radiation pressure plays in the outflow launching process~\citep{pereira-santaella_are_2021}. 
 Most of the outflowing gas mass is loaded from within the central few hundred parsec of the galactic cores. 
 Outflow depletion timescales of $< 10^8$ years 
 are found to be anti-correlated with AGN luminosity and much shorter than the gas-consumption-by-star-formation timescales~\citep{gonzalez-alfonso_molecular_2017,gowardhan_dual_2018,pereira-santaella_spatially_2018}. 
 
 In a study of near-infrared \molhy~in 21 GOAL (U)LIRGs, the shocked molecular gas is preferentially found in the ultraluminous systems but is also seen to be triggered at a lower-luminosity, earlier merging stage~\citep{u_keck_2019}. 
 In a separate study of CO and \molhy~in 6 (U)LIRGs with Seyfert nuclei, those galaxies undergoing major mergers are found to exhibit shallower temperature distributions and, thus, a higher relative contribution of shock excitation with respect to the photodissociation region, for the warm molecular gas than the normal spirals~\citep{Pereira-Santaella14}. 
 These findings raise the question of whether AGN---also commonly found in the more luminous systems---are responsible for shock-heating the gas. On circumnuclear scales, AG nuclei have strong effects on heating the molecular gas, but the shocks do not strictly correlate with AGN strength. Such a link may be muddled by the presence of patchy dust, torus orientation, varying gas density, or other factors~\citep{u_keck_2019}.

\subsubsection{The AGN--Outflow--Merger--ULIRG Connection}
At the population level, multiphase winds occur in almost all (U)LIRGs with SFR above 10 $M_\odot$ year$^{-1}$~\citep{Rupke05a,Rupke05b,Rupke05c}. How much of this wind is powered by AGN versus the underlying starburst population requires detailed investigations of the wind geometry (i.e., opening angles) as well as momentum and energy injection into the ISM among Seyfert 1, Seyfert 2, and starburst-dominated ULIRG systems~\citep{Rupke05d}. In a systematic search for OH molecular outflows in 43 nearby ultraluminous infrared galaxy mergers,~\citet{Veilleux13} measured higher outflow velocities in systems with large AGN fractions and high AGN luminosities [$\log (L_\mathrm{AGN}/L_\odot) \geq 11.8 \pm 0.3$]. However, the most AGN-dominated systems having relatively modest OH line widths might suggest an evolutionary scenario where molecular outflows subside at the quasar blow-out phase. 

\begin{figure}[H]

\includegraphics[width=.9\textwidth]{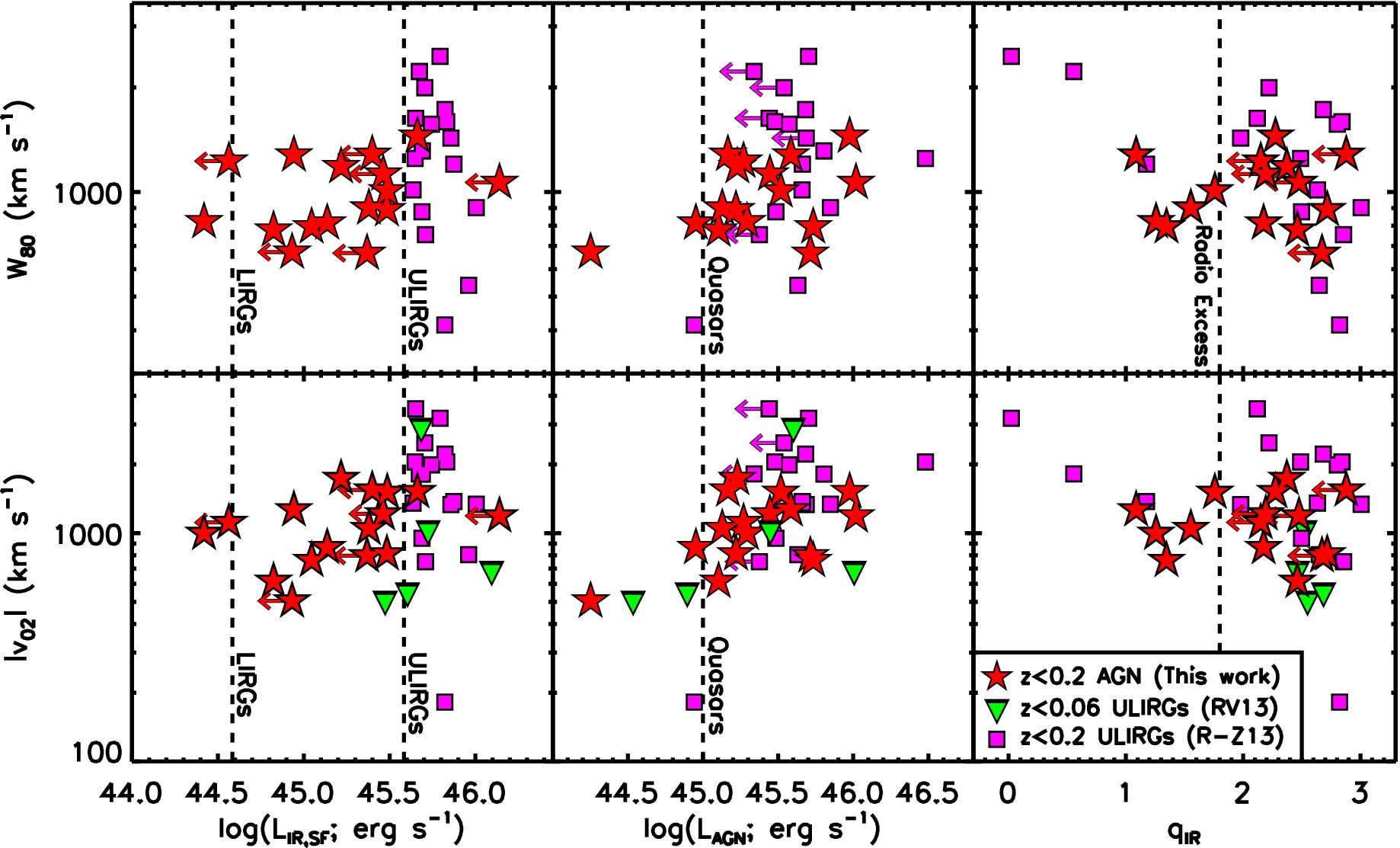}
\caption{Outflow velocity vs. infrared luminosity from star formation (\textbf{left}), AGN luminosity (\textbf{middle}), and $q_\mathrm{IR}$ (\textbf{right}) for nearby AGN~(red stars; \cite{harrison_kiloparsec-scale_2014}), ULIRGs~(green triangles; \cite{Rupke13}), and ULIRG-AGN composites~(magenta squares; \cite{Rodriguez-Zaurin13}). The most extreme velocities appear to be preferentially found in ULIRGs that host quasars. Figure adapted from \citet{harrison_kiloparsec-scale_2014} with permission.
\label{fig:harrison}} 
\end{figure}

The scenario that AGN are more likely to be buried, or less luminous, during late-stage mergers associated with the highest molecular outflow velocities and short gas depletion timescales seem contradictory to studies that found the fastest ionized winds in the most luminous AGN. For instance, \citet{harrison_kiloparsec-scale_2014} compared the velocities of ionized outflows in AGN, ULIRGs, and ULIRG-AGN composites and found that the most extreme ionized gas velocities ($v \gtrsim 2000$ km s$^{-1}$) are preferentially found in quasar-ULIRG composite galaxies (Figure~\ref{fig:harrison}~\citep{harrison_kiloparsec-scale_2014}). This is likely because AGN and highest-SFR starbursts are confounded in late-stage merging ULIRGs, where distinguishing between the two as the origin of the outflows is non-trivial. In this case, the starburst may contribute a non-negligible (if not dominant) amount to the ionized outflows seen at the kiloparsec scale. 
A recent study of highly-ionized \ovi~and \nv~outflows in 19 ULIRGs found that outflows are more frequently detected in the X-ray weak or absorbed sources, which also feature higher-velocity outflows~\citep{Liu22}.
The seemingly contentious point rests on whether ULIRGs host deeply buried AGN or luminous quasars, and may be reconciled if the obscuration timescales and properties in ULIRGs can be better quantified with imminently available high-resolution \emph{JWST} observations for large samples of local ULIRGs.

\section{Dual SMBHs in Merger Systems}
\label{sec:duals}

\subsection{The Observational Challenges and Significance of Constraining the SMBH Pair Population}
\begin{figure}[H]

\includegraphics[width=.65\textwidth]{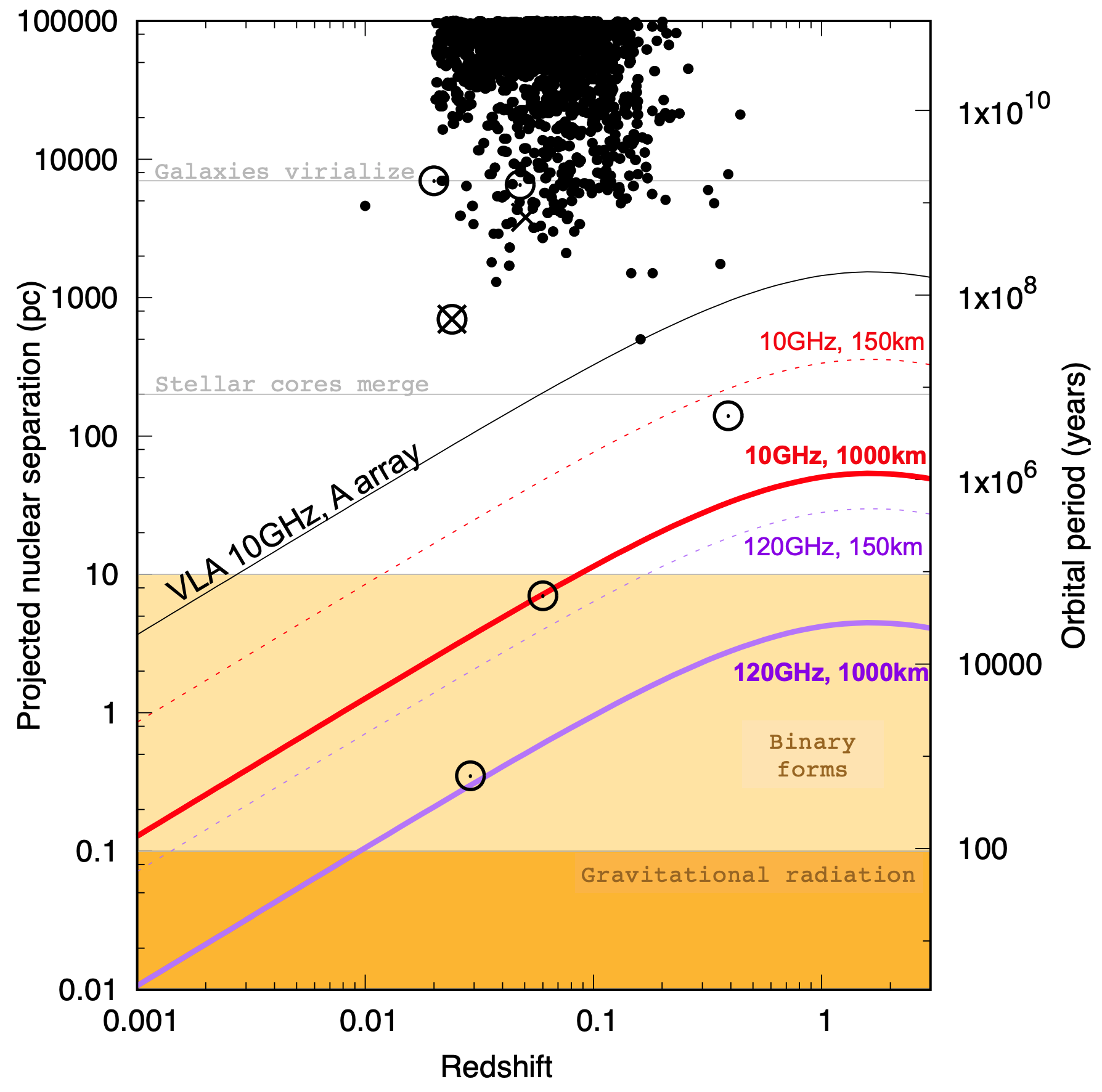}
\caption{A compilation 
of candidate AGN pairs to date shown in projected nuclear separation---redshift space. The dual AGN are indicated by their discovery method: optical/near-infrared (dots, from~\citet{Liu11}), X-ray (circles), and radio (crosses). The colored dashed and thick curves represent the resolution limits of the ngVLA at different frequency--baseline configurations with the VLA (10 GHz, A array; thin black line) as a comparison. Critical stages in binary formation and evolution are marked as horizontal lines. Only a handful of duals and binaries have been identified with nuclear separation below 1 kpc, so a key area of research would be to locate these SMBH pairs in mergers and post-mergers using new multiwavelength facilities with enhanced resolving power in the coming decade.
Figure adapted from \citet{Burke-Spolaor18} with permission.} 
\label{fig:burke-spolaor} 
\end{figure}

In the era of multi-messenger astrophysics, the science case for studying merging SMBHs is prominent given the anticipated detection of gravitational wave signatures from this massive regime. Understanding the statistical properties of the merging SMBH population or the nuclear conditions of the progenitor galaxies has great implications for the expected detection rate of gravitational wave signals. Naturally, close SMBH pairs are expected to be found within the circumnuclear disks in late-stage or post-mergers. A large number of dual AGN candidates with nuclear separations of 1 kpc or greater have been discovered using optical methods~\citep{Liu11}, but the $r_\mathrm{sep} < 1$ kpc regime has presented many more challenges. While several binary SMBH candidates 
have been \mbox{posited~\citep{Fu11,kharb_candidate_2017,Dey18}}, only one appears robust to date (with $r_\mathrm{sep} = 7.3$ pc in between the two SMBHs as resolved by VLBI~\citep{Rodriguez06}). A compilation of candidate AGN pairs discovered is summarized in Figure~\ref{fig:burke-spolaor}.

Observational challenges to finding merging SMBHs are several-fold: scientifically, close SMBH pairs might be missed from traditional optical searches if they reside in the dust-obscured nuclei of late-stage mergers, where intense starbursts and other sources of extreme energetics might contribute to the contamination rate of AGN detection. The two AGN may also engage in asynchronous triggering, causing the quiescent black hole in the pair to be off the radar. On the technical front,
the best-case resolving capability---the diffraction limit $\theta$---scales with the aperture size $D$ of an observing facility for a given wavelength  of interest $\lambda$ in the form of $\theta = 1.22 \lambda/D$.
Specific detection methods may also introduce biases and blind spots.
Hence, the multi-messenger search of true inspiraling SMBH binaries prompts multiple validations, for instance, using a combination of gravitational wave detectors, such as the pulsar timing arrays (PTAs~\citep{Foster90,Hobbs10}) 
and large time-domain surveys providing multi-epoch light curves to verify the periodicity of the merging pair.

\textls[-15]{In the massive black hole binary regime, observational evidence for identification includes the following: emission of gravitational waves, direct imaging of double nuclei using VLBI, quasi-periodicity in light curves of AGN and quasars, and double-peaked emission lines in AGN and quasar spectra.
For more details, we refer the readers to articles dedicated to the search for binary SMBHs by~\citet{Burke-Spolaor19,deRosa19,Bogdanovic21}.}
In the dual regime with larger nuclear separation, additional identification and verification techniques may include determining the dynamics around black holes and directly resolving hard X-ray sources given the larger separation allowance. The key is to be able to 1) resolve the two (or more) point sources within the distance limitation, and 2) be able to unambiguously confirm the SMBH nature for each of the point sources at the wavelength(s) of interest. 

Our discussion here highlights observational results of dual SMBHs with projected separation $r_\mathrm{sep} \approx 10 - 1000 $ pc residing in the center of merging infrared galaxies. The duals are scientifically relevant because they are the precursors to the gravitationally-bound binaries and an understanding of the environment in which they are located will give us insights into the dual AGN discoverability and the dynamical timescale at the population level. The amount of infalling gas into the center triggered by the merger's orbital dynamics may have huge implications on whether the binary SMBHs will eventually coalesce and produce gravitational wave signals or if they will instead be stalled (e.g., the ``final parsec problem''~\citep{Milosavljevic03}). 


\subsection{The Multiwavelength Searches for Duals in Mergers}

Much of the search for double AGN has taken advantage of the high spatial resolution capability of interferometry in the radio regime. Thanks to their long baseline configurations, extended radio interferometry arrays, such as the Very Long Baseline Interferometry (VLBI), can probe down to 1 $\times$ $10^{-6}$ ArcSec~\citep{Bartel88}, or equivalently, the sub-parsec regime in local sources. Given the power of the radio spectral index to detect high brightness temperature and synchrotron emission from AGN in even dusty nuclei (see Section~\ref{sec:agn}), a technique using VLBI to resolve and identify multiple point sources promises to be very feasible. Such a blind search was conducted using archival VLBI data of radio-luminous AGN~\citep{burke-spolaor_radio_2011}. However, only one binary black hole candidate was confirmed among 3114 sources examined: 0402+370, or 4C +37.11, which hosts two nuclei separated at 7 pc as previously discovered by~\citet{Rodriguez06}. Taking into account the limitations in the spatial resolution of the large redshift range ($z = 0.000113 - 4.715$, where projected separation from sub-parsec up to 8.5 pc can be resolved) probed in the sample, the implications from this statistical result place constraints on the physics of merging SMBH pairs and the broader cosmological context of the merger population. Whether both binary SMBHs are unlikely to be found in a radioactive state simultaneously or if the timescale for an inspiraling SMBH pair to have $<$10 pc separation is very short will require further investigations, possibly with the future square kilometer array. 

Optical long-slit spectroscopy can be useful in helping to verify or rule out the nature of dual AGN given the diagnostic power of BPT or equivalent line ratio diagrams (see Section~\ref{sec:agn}). 
Dual AGN selection using the presence of a double-peaked \oiii~$\lambda$5007 narrow emission line has been favored given the abundance of rest-frame optical spectra of galaxies and quasars available through the Sloan Digital Sky Survey (SDSS) and other large spectroscopic surveys. A double-peaked \oiii~line may originate from the two distinct NLRs surrounding the two black holes in a dual or binary system~\citep{Boroson09,Wang09,Comerford09,Comerford13,McGurk11,Barrows12,Fu12}, but it may also be produced by biconical or one-sided outflows in a single black \mbox{hole \citep{Veilleux01,Fischer11,Rupke13,Liu18}} or a rotating disk of NLR associated with a single black hole~\citep{Greene05,Liu10,Smith12}. Additional scrutiny is needed to vet the contaminants. 
Using either IFU or a combination of high-resolution imaging plus resolved spectroscopy, \citet{McGurk15} studied a sample of 140 SDSS DR7 sources with double-peaked narrow \oiii~lines and found 30\% having two spatial components. A follow-up in-depth study of 12 candidates within this subset substantiated only one dual AGN system (at separation 6.6 kpc) with additional outflow or other candidates~\citep{McGurk15}. A larger, more recent study of 1271 SDSS quasars at $z < 0.25$ with multiwavelength verification methods found a similar success rate (77 dual candidates, or 6\%)~\citep{Kim20}.

\begin{figure}[H]

\includegraphics[width=.65\textwidth,trim=0 480 0 0,clip]{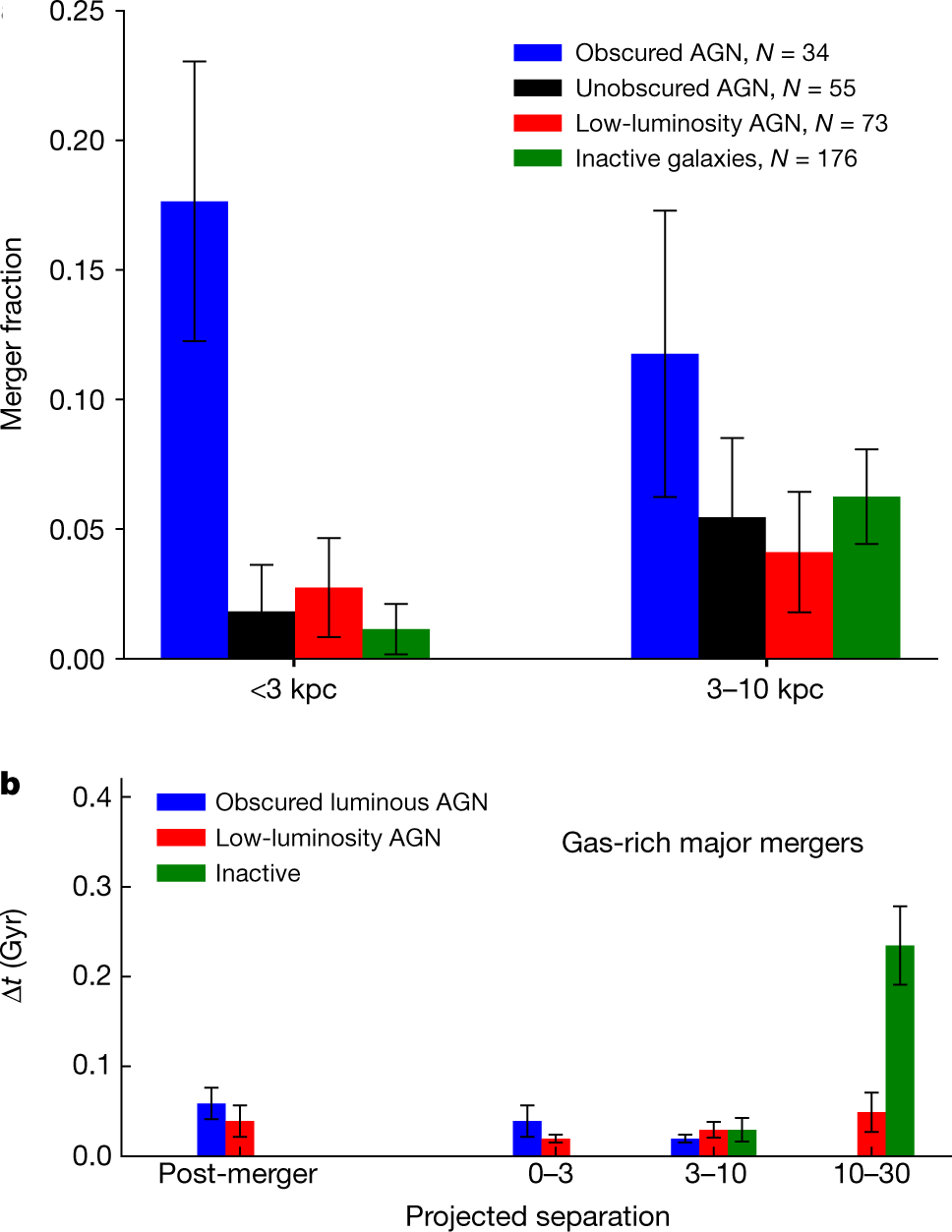}
\caption{Distributions 
of close mergers at the $<$3 kpc and 3--10 kpc separations, respectively, as categorized by the presence of broad H recombination lines from optical spectroscopy and bolometric luminosity as computed from X-ray emission. There is clearly an excess of nuclear mergers among obscured AGN at close separations. Figure adapted from \citet{koss_population_2018} with permission.
\label{fig:koss}} 
\end{figure}

While hard X-ray observations provide a robust tool for locating highly obscured sources that are most likely to be close SMBH pairs during the late stages of mergers, current X-ray facilities with photon-counting techniques tend not to have high enough spatial resolution to resolve close pairs at less than 0\farcs5 separation, though deep X-ray observations of putative duals can help confirm their candidacy~\citep{Comerford15}. Combining a hard X-ray selection with systematic high-resolution AO-assisted near-infrared imaging of 96 low-redshift (0.01 $< z <$ 0.075) Swift/BAT-detected sources, \citet{koss_population_2018} confirmed the excess of nuclear obscured duals with separations $<$ 3 kpc among late-stage mergers (17.6\%) compared to a control sample of inactive galaxies (1.1\%), consistent with theoretical predictions (Figure~\ref{fig:koss}). Indeed, another X-ray + near-infrared spectra + \emph{WISE} study of advanced mergers with projected separations $<$ 10 kpc confirmed that optical studies alone miss a significant fraction of single and dual AGN in advanced mergers, advocating that mid-infrared pre-selection can highly enhance the success rate of identifying dual AGN~\citep{Satyapal17,Pfeifle19}.

A general note of caution is that resolving multiple point sources is a necessary but insufficient criterion for identifying dual AGN, particularly in galaxy mergers where the central regions tend to be complex due to the presence of dust, nuclear star clusters, and the corresponding entangled dynamics. Sometimes it might be difficult to define where the ``center'' of an interacting system is, such as in II Zw 96, where the most infrared bright source appears off-nuclear~\citep{Inami10}). 
\citet{medling_stellar_2014} presented the nuclear stellar and gas kinematics for a sample of 17 (U)LIRGs at a high ($\leq$ 0\farcs1) 
resolution: dynamically-rotating stellar and gaseous disks are nearly ubiquitous, but in most cases, nuclei with resolved point sources do not display a superposition of two (or more) dynamical disks. While not completely ruled out, those sources are more likely to host a remnant nucleus with a nearby star cluster that appears close-by by chance projection than a circumbinary disk with merging black holes.


\subsection{Implications from Duals in Merging (U)LIRGs}
The current list of confirmed dual AGN with $r_\mathrm{sep} \lesssim 1$ kpc among local (U)LIRGs consists of only a handful of sources: NGC 6240 (1 kpc;~\citep{Komossa03}), Arp 299 (900 pc;~\citep{Ballo04}), and Mrk 273 (800 pc;~\citep{iwasawa_location_2011,u_inner_2013,liu_elliptical_2019}). (U)LIRGs that feature dual AGN with slightly larger separations include Mrk 266 (6 kpc;~\citep{Mazzarella12}), Mrk 463 (3.8 kpc;~\citep{Bianchi08,treister_optical_2018,Yamada18}, and borderline-LIRG Mrk 739 (3.4 kpc;~\citep{Koss11}). NGC 7674, a moderately-luminous LIRG with a companion 20~kpc away from the main infrared emission source, was identified as a binary SMBH host candidate within its nucleus based on VLBI results~\citep{kharb_candidate_2017}, but the signal-to-noise ratio of the radio imaging is marginal and its candidacy was more recently challenged~\citep{Burke-Spolaor19}.  

\begin{figure}[H]

\includegraphics[width=.6\textwidth]{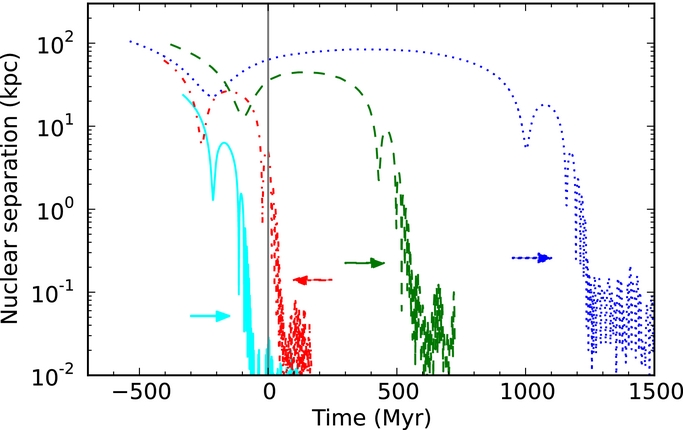}\\
\includegraphics[width=.6\textwidth]{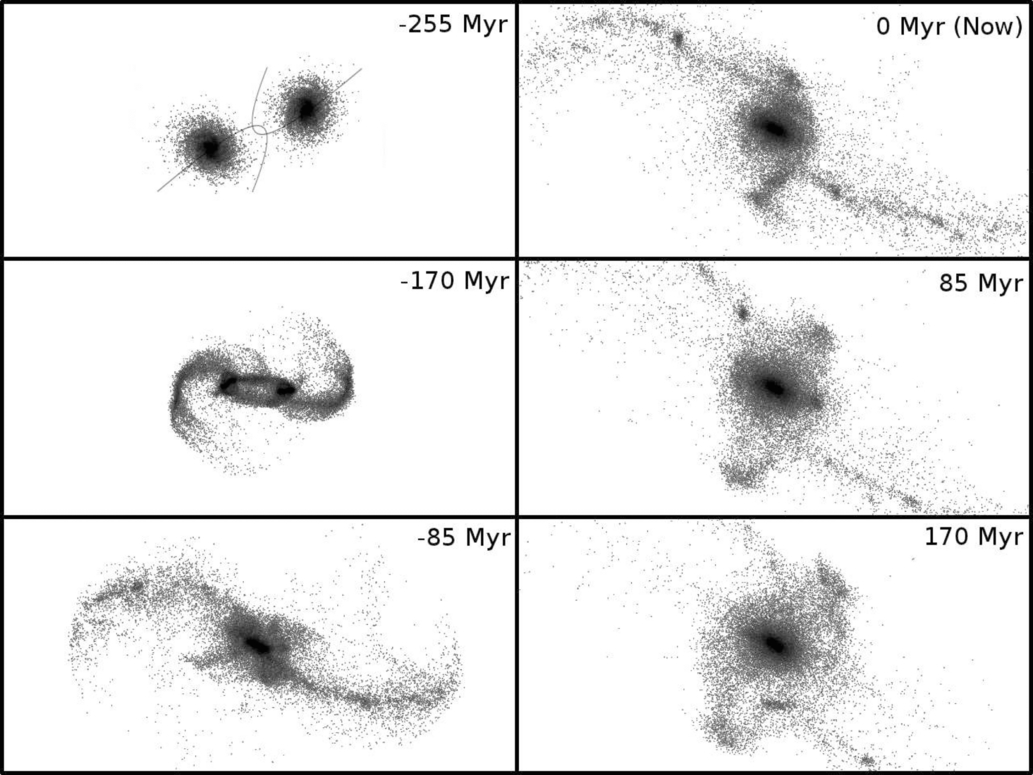}
\caption{(\textbf{Top}) Nuclear separation as a function of time for four merging systems (from left to right: NGC 2623, Antennae, The Mice, and NGC 5257/8) as determined from Identikit~\citep{Barnes09} dynamical modeling of H\textsc{i} and optical imaging. The vertical line at the time of zero represents the current viewing time. Colored arrows mark the corresponding smoothing length in kpc for each system, below which the spatial resolution of the simulations may not be reliable. (\textbf{Bottom}) Example snapshots of how LIRG NGC 2623 evolved as a function of time. The grey lines show the initial Keplerian orbits of the galaxies, and the times shown are relative to now. Figures adapted from \citet{Privon13};~\textcopyright~AAS, reproduced with permission.
\label{fig:privon}} 
\end{figure}

Nevertheless, the identification of close SMBH pairs in these (U)LIRGs suggests that they might represent the best bet for locating progenitor merging SMBHs in order to understand the small-scale environments that favor pairs. Ongoing galaxy mergers may preferentially be found in group environments where galaxies are more susceptible to interacting with one or more partner(s) in the recent past, so characterizing the SMBH-pair rate in these systems at the population level could place constraints on our understanding of the evolution of the merger fraction or the timescale of synchronous AGN-triggering. Detailed dynamical modeling of merging systems (Figure~\ref{fig:privon};~\citep{Privon13}) coupled with the detection of dual SMBHs could address these questions, though accomplishing this task at the population level could be resource intensive.

The conundrum that (U)LIRGs in the local universe are relatively rare and yet our current state-of-the-art resolving capability limits our electromagnetic search for duals and binaries to the local universe means that our next breakthrough will be instigated by the next-generation observing facilities. 
More cutting-edge technology across the electromagnetic spectrum pushing the angular resolution limit will hopefully render the search for SMBH merger progenitors more fruitful within the next decades.
With the onset of PTAs and other next-generation gravitational wave detectors, such as LISA, prompted to detect gravitational wave signals from the merging of SMBHs, it is more timely than ever that we investigate the nuclear environments within (U)LIRGs that lead to black hole mergers.

\section{Future Outlook from the Observational Perspective}
\label{sec:future}

In this review, we covered recent observational results in identifying AGN via a multiwavelength approach, investigating AGN fueling and feedback from the galactic core to halo, and locating close SMBH pairs within merging infrared galaxies. These science cases align primarily with one of the three priority science themes identified by the Astro2020 Decadal Survey---understanding Cosmic Ecosystems through unveiling the drivers of galaxy growth and exploring the interconnectedness of different processes responsible for galaxy evolution. A large part of this multiscale problem lies in understanding the physics of AGN, driving investments in facilities that provide high spatial resolution. Here, we discuss how several of the upcoming observatories and facilities with unparalleled technical capabilities will help achieve breakthrough science cases in our understanding of AGN's role in infrared galaxies.

\subsection{Peering through the Dust}
The \emph{JWST} just started science operations at the time of this manuscript's preparation. 
As a NASA flagship, \emph{JWST} is expected to extend the legacy of both \emph{HST} and \emph{Spitzer} as a space telescope with near- and mid-infrared (0.8--28~$\upmu$m) wavelength coverage. Its mid-infrared wavelength reach coupled with its 6.5-m mirror will surpass \emph{Spitzer}'s sensitivity and spatial resolution (with a 50$\times$ smaller PSF area), enabling observations to sample the rest frame UV and optical light from high-redshift galaxies and probes far into the early universe. To facilitate the search for CT AGN, the mid-infra-red instrument (MIRI) will be able to observe the forbidden neon lines with ionization potentials (IPs) ranging from 21.6 (\neii) to 126 eV (\nevi)~\citep{Alberts20} or the highly-ionized \mgiv~$\lambda$4.487 $\upmu$m~(IP of 80 eV) and \arvi~$\lambda$4.528 $\upmu$m~(IP of 75 eV)~\citep{Satyapal21}. 
These lines are associated with the extra hard-ionizing radiation from the AGN given their high ionization potentials. For instance, the enhanced \neiii~$\lambda$15.56 $\upmu$m/\neii~$\lambda$12.81 $\upmu$m~ratio measured in three Seyfert galaxies relative to the median value of that for a LIRG sample showing no AGN signatures demonstrates their utility as robust AGN indicators~\citep{Pereira-Santaella10}.
MIRI MRS has the potential to acquire mid-infrared spectra of (U)LIRGs up to $z \sim 1-2$ that includes \nevi~$\lambda$7.65 $\upmu$m. The synergy of the MIRI imager and NIRCam will enable us to detect heavily obscured and CT AGN while possibly associating them to mergers at high redshifts. See \mbox{\citet{Alberts20}} and \citet{Satyapal21} for more quantitative discussions.

\begin{figure}[H]

\includegraphics[width=.6\textwidth]{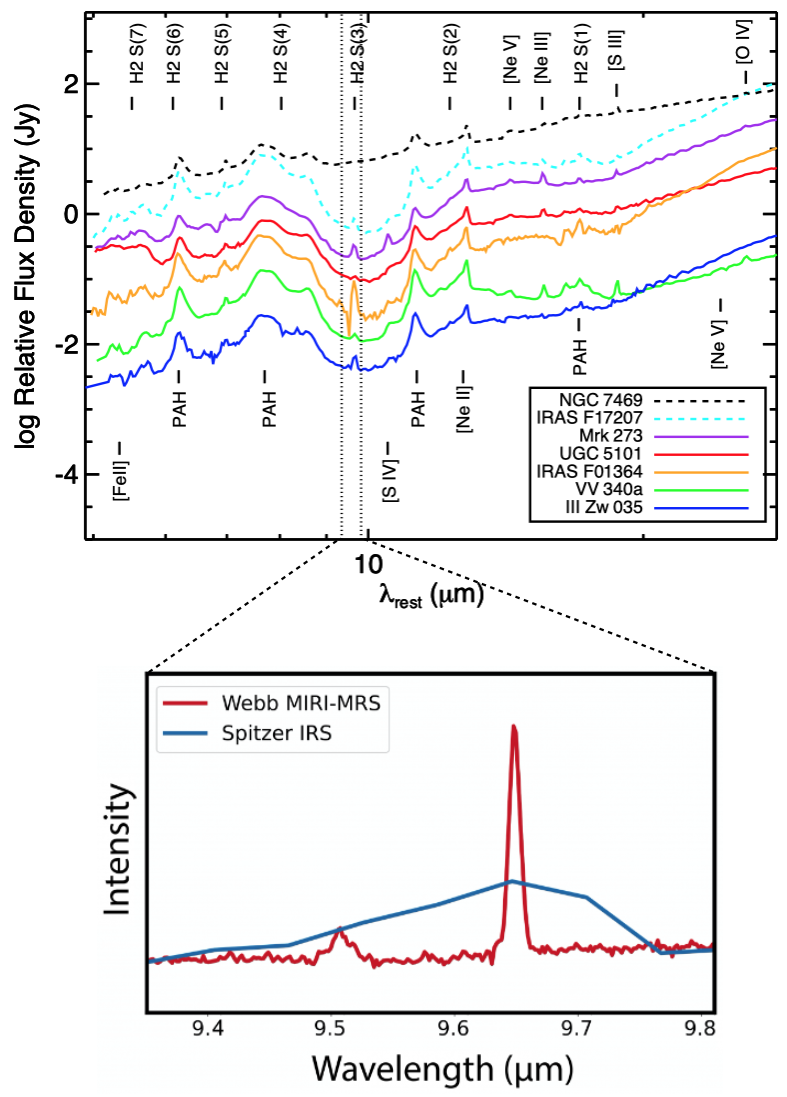}
\caption{\emph{Spitzer}/IRS spectra for several (U)LIRGs in GOALS~\citep{stierwalt_mid-infrared_2013}. A full suite of crucial mid-infrared diagnostic lines are accessible and will probe excitation mechanisms, extincted star formation, dust and metal content, and AGN strength and ionization state in typically heavily obscured regions otherwise opaque to optical studies. The spatial and spectral resolutions of the IRS spectra (R$\sim$60--600) will be superseded by MIRI MRS (R$\sim$2500). Dotted vertical lines indicate the spectral region covered by the zoomed-in inset. (Inset) New MIRI MRS engineering calibration spectrum of Seyfert galaxy NGC 6552 (red), shown in comparison with a lower spectral resolution IRS spectrum of a similar galaxy in the same spectral region (blue). The strong emission feature at rest frame 9.66~$\upmu$m~comes from molecular hydrogen \molhy~S(3), while the weaker feature is likely \fevii~$\lambda$9.53~$\upmu$m.  Inset figure from NASA, ESA, and the MIRI Consortium (courtesy of David Law and Alvaro Labiano). 
\label{fig:irsa}} 
\end{figure}

As for nearby infrared galaxies, early release science, guaranteed time observations, and general observer programs in Cycle 1 will capitalize on MIRI being the first mid-infrared IFU in space to investigate the kinematics of gas and dust in the dusty cores of (U)LIRGs. With a pixel scale of $\sim$ 0\farcs2 pixel$^{-1}$, MIRI in MRS mode is well-suited to study the resolved morphology of the cool ($\sim$1000 K) molecular gas, silicate absorption, and PAH features often found in (U)LIRGs (Figure~\ref{fig:irsa}).  MIRI's imaging mode has a plate scale of 0\farcs11 pixel$^{-1}$, fully sampling the JWST point spread function at 5.6 $\upmu$m. It will be able to image the AGN dusty torus at $<$10 pc resolutions within the Seyfert nuclei of the nearest AGN, such as LIRG NGC 1068.  

\begin{figure}[H]

\includegraphics[width=.8\textwidth]{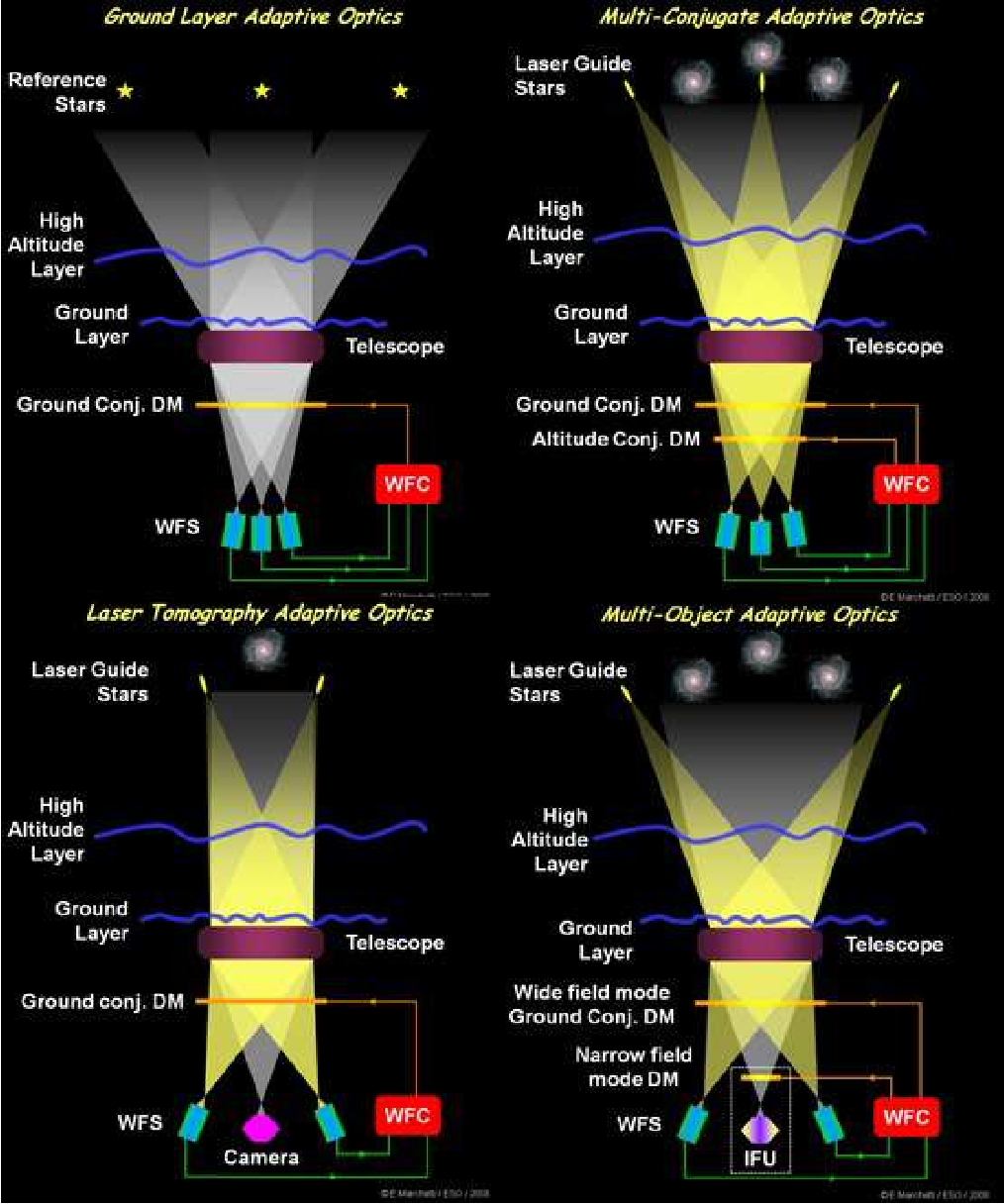}
\caption{\textls[-25]{Conceptual illustrations of several AO techniques that aim to improve the sky coverage, Strehl ratio, or image quality of ground-based observations. Figure adapted from \citet{Davies12} with permission}. 
\label{fig:ao}}
\end{figure}

In the near-infrared regime, \emph{JWST} excels in sensitivity given its large 6.5-m mirror in space, while the angular resolution afforded by its near-infrared camera NIRCam (0\farcs031--0\farcs063 pixel$^{-1}$) and IFU NIRSpec (0\farcs1 pixel$^{-1}$) is matched by diffraction-limited AO observations from large, 8- to 10-m class ground-based facilities. The Keck Telescopes, for instance, host AO-assisted near-infrared IFU (OSIRIS) and camera (NIRC2) whose smallest plate scales are 0\farcs020 and 0\farcs01, respectively, with a typical FWHM of 0\farcs05 under good seeing conditions. These observations have been instrumental in mapping the nuclear dynamics surrounding the AGN in nearby (U)LIRGs and finding dual AGN within the dusty cores (see Sections \ref{sec:feedback} and \ref{sec:duals} for examples).  
Existing natural and laser guide star AO systems have been phenomenal in resolved galactic nuclei studies, and novel techniques (such as multi-conjugate AO) that allow multiplexing, ground-layer, and all-sky AO that strive to improve sky coverage, Strehl ratio, or image quality, will greatly enhance the number of galaxy systems that can be targeted from the ground to broaden the demographics (Figure~\ref{fig:ao}; see~\citet{Davies12} for an in-depth discussion of different AO techniques). 

\subsection{A Leap in Resolution}
The next step in technological advances is to push the frontier of AO diffraction-limited observations to the visible wavelength range. 
This goal is currently being realized by an ongoing mission named ORCAS, or the Orbiting Configurable Artificial Star. As a hybrid space and ground observatory, the ORCAS mission\endnote{\url{https://asd.gsfc.nasa.gov/orcas/about/} (accessed on 24 June 2022)} 
(PI E. Peretz) beams down a laser to the ground-based observatory and provides AO corrections and flux calibrations over the wavelength range 0.5--2 $\upmu$m. Coupled with the 10-m mirror at Keck Observatory, ORCAS will enable near-diffraction limited performance (high angular resolution ($\sim$12--40 mas); high spectral resolution (R $\lesssim$ 10,000); and high sensitivity (29th magnitude)) from the ground at visible wavelengths, detecting AGN feeding and feedback signatures within the inner pc-scale regions and resolving SMBH binaries in the near universe. At its best, ORCAS$+$Keck will be able to resolve close SMBH pairs with separations down to $\sim$100 pc up to redshift $z \sim 2$, where the (U)LIRG population dominates the star formation rate density~\citep{Casey14}. The visible-light imager ORCAS-Keck Instrument Demonstration (ORKID; currently being commissioned in Semester 2022B), together with a near-diffraction-limited visible IFU in the works, will place critical constraints on the number of optically-selected dual AGN pairs among SMBHs and measure the merger fraction throughout cosmic history.

\begin{figure}[H]

\includegraphics[width=.7\textwidth]{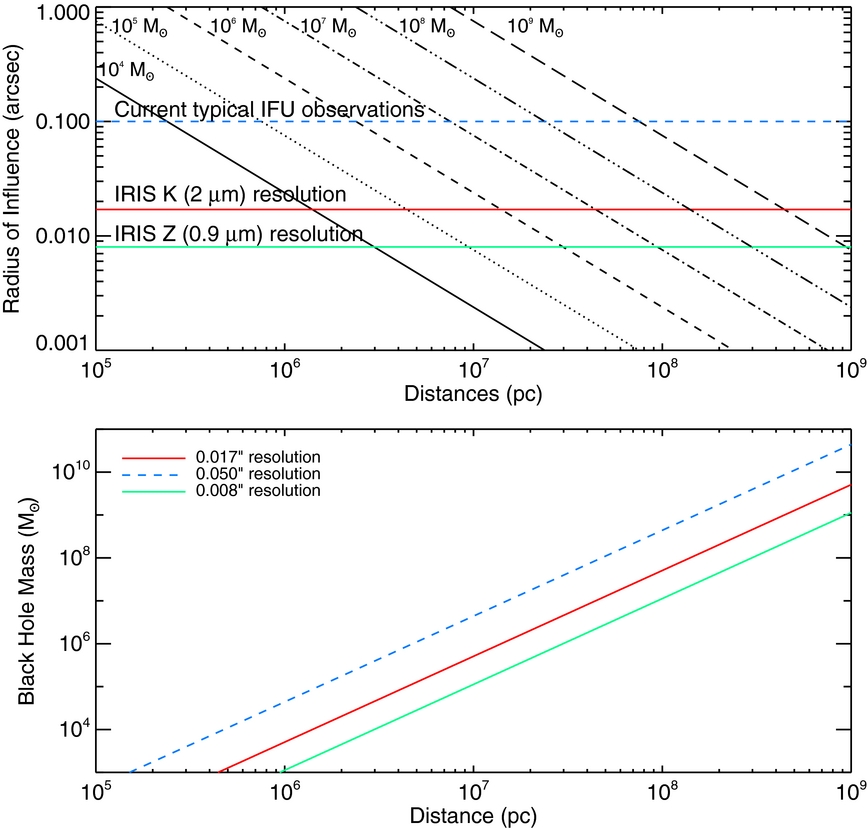}
\caption{(\textbf{Top}) The projected radius of influence for a range of black hole masses (black lines) 
plotted as a function of distance. Overlaid are the angular resolution limits of current typical IFU observations (blue dashed) and those anticipated for IRIS at $K$ (red) and $Z$ (green) bands, respectively. The enhancement in resolving capability brought about by 30-m class telescopes will push the frontier of black hole mass measurements to further distances. (\textbf{Bottom}) Lower limits on the black hole mass measurements are shown as a function of distance. Figure from Do et~al.~\cite{Do14};~\textcopyright~AAS, reproduced with permission. 
\label{fig:do_tmt}}
\end{figure}

The end of the 2020s and the 2030s will welcome the 30-m class extremely large telescopes (ELTs), including the European-ELT (E-ELT), the Giant Magellan Telescope (GMT), and the Thirty-Meter Telescope (TMT). These ELTs will be revolutionary in providing optical and near-infrared observations from the ground at superb sensitivity and resolution. Coupled with the large glass, advanced IFU instruments, such as the high angular resolution monolithic optical and near-infrared integral-field spectrograph (HARMONI) on the ELT or the infra-red imaging spectrograph (IRIS) on the TMT will reach resolutions of 0\farcs009 
at $\sim$1--2 $\upmu$m, ideal for probing down to the dusty tori around AGN. Figure~\ref{fig:do_tmt} shows the capability of TMT/IRIS in resolving the sphere of influence for black holes of certain masses at a range of distances. The IFU ability to determine kinematics around black holes will allow not just black hole mass determination, but also putative dual AGN identification as well as a direct probe into the feeding and feedback mechanisms around one or more AGN in infrared galaxies beyond the local universe.

\section{Concluding Remarks}
\label{sec:conclusions}

In this review, we presented recent multiwavelength observational results related to the detection of one (or more AGN) in nearby (U)LIRGs, which are mostly gas-rich major mergers, in addition to the fueling and feedback roles they play in regulating galaxy growth. Unlike using homogeneous quasars or Seyfert samples, (U)LIRG studies present additional challenges in finding the true location and verifying the nature of the nuclei within merging systems, thanks both to the galaxy interaction and dust obscuration. Detailed understanding of the nuclear regions requires high spatial resolution observations, which are limited to the local universe where there are relatively rare (U)LIRGs. Nevertheless, we as a community have built up statistical samples that altogether begin to shed light on (U)LIRGs' place within galaxy evolution. Upcoming space- and ground-based facilities will provide a leap in resolution across the electromagnetic spectrum and allow us to expand our investigations both deeper within the galactic nuclei and broader to a larger population at higher redshifts. The coming decade will be a golden age for studying AGN in infrared galaxies.

%

\vspace{6pt} 




\funding{This research was funded by National Aeronautics and Space Administration Astrophysics Data Analysis Program grant number 80NSSC20K0450.}

\acknowledgments{V.U. thanks A. Sajina, G. Rieke, S. Yamada, L. Barcos-Mu\~noz, G. Privon, A. Medling, J. Mazzarella, D. Sanders, A. Stacey, A.\,Evans,\,and\,the\,two\,anonymous\,referees for insightful comments that contributed to the framing and content of this review.  
This research has made use of the NASA/IPAC Extragalactic Database, which is funded by the National Aeronautics and Space Administration and operated by the California Institute of Technology.
} 

\conflictsofinterest{The funders had no role in the design of the study; in the collection, analyses, or interpretation of data; in the writing of the manuscript, or in the decision to publish the~results.} 



\abbreviations{Abbreviations}{
The following abbreviations are used in this manuscript:\\

\noindent 
\begin{tabular}{@{}ll}
AGN & Active Galactic Nuclei\\
ALMA & Atacama Large Millimeter/submillimeter Array \\
AO & Adaptive Optics\\
BAT & Burst Alert Telescope\\
CON & Compact Obscured Nuclei\\
CT & Compton Thick\\  
ELT & Extremely Large Telescope\\
GOALS & Great Observatories All-sky LIRGs Survey\\
HST & Hubble Space Telescope\\
IFU(S) & Integral Field Unit (Spectroscopy)\\
IRAS & Infrared Astronomical Satellite\\
ISM & Interstellar Medium\\
IP & Ionization Potential\\
JWST & James Webb Space Telescope\\
LIRG & Luminous Infrared Galaxy\\
MIRI & Mid-Infra-Red Instrument \\
MRS & Medium Resolution Spectroscopy \\
MUSE & Multi Unit Spectroscopic Explorer \\
NIRCam & Near Infrared Camera \\
NIRSpec & Near Infrared Spectrograph \\
NLR & Narrow Line Region\\
\textbf{PG} & \textbf{Palomar-Green (quasar)} \\
PUMA & Physics of ULIRGs with MUSE and ALMA \\
PTA & Pulsar Timing Arrays \\
QSO & Quasi Stellar Object \\
QUEST & Quasars/ULIRGs Evolutionary STudy\\
SDSS & Sloan Digital Sky Survey\\
SFR & Star Formation Rate\\
SFMS & Star Formation Main Sequence\\
SMBH & Supermassive Black Hole\\
ULIRG & Ultraluminous Infrared Galaxy\\
VLA & Very Large Array\\
VLBI & Very Long Baseline Interferometry\\
WISE & Wide-field Infrared Survey Explorer (WiSE)
\end{tabular}}




\begin{adjustwidth}{-\extralength}{0cm}
\printendnotes[custom]
\newpage
\reftitle{References}



\end{adjustwidth}

\end{document}